\documentclass[11pt]{article}
\pdfoutput=1

\usepackage{etexcmds}
\usepackage{jcapmod}
\usepackage{booktabs}
\usepackage[english]{babel}
\usepackage{amsmath,amssymb,amsbsy,amstext, amsthm, simplewick}
\usepackage{graphicx}
\usepackage{amsfonts}
\usepackage{amssymb}

\usepackage{upgreek}
\usepackage{exscale,relsize}
\usepackage[makeroom]{cancel}
\usepackage{soul}
\usepackage{adjustbox}
\usepackage{slashed}
\usepackage[font=footnotesize]{caption}

\usepackage{colortbl}
\definecolor{rp}{cmyk}{0.2, 1, 0.6, 0}
\definecolor{green2}{cmyk}{0, 1, 0.5, 0}
\definecolor{lightgreen}{cmyk}{0.2, 0, 0.2, 0.2}
\definecolor{lightgray}{cmyk}{0.1,0.2,0,0.1}
\definecolor{lightgray2}{cmyk}{0.4,0.4,0,0.8}
\definecolor{black}{cmyk}{1.0,1.0,1.0,1.0}

\numberwithin{equation}{section}

\definecolor{MyBlue}{rgb}{0.15,0.15,0.70}

\hypersetup{
colorlinks=true,
citecolor=MyBlue,
linkcolor=MyBlue,
urlcolor=MyBlue
}

\input epsf
\newcommand{\project}[1]{{\sffamily #1}}

\DeclareMathOperator{\Tr}{Tr}

\newcommand{\be}{\begin{equation}}
\newcommand{\ee}{\end{equation}}
\newcommand{\beq}{\begin{equation}}
\newcommand{\eeq}{\end{equation}}
\newcommand{\bea}{\begin{eqnarray}}
\newcommand{\eea}{\end{eqnarray}}

\newcommand{\bfnl}{b_{\rm NG}}

\renewcommand\[{\left[}

\def\k{{\bf k}}
\def\q{{\bf q}}
\def\p{{\bf p}}

\def\btheta{{\boldsymbol \theta}}
\newcommand\ees{\end{eqnarray}}
\newcommand\bees{\begin{eqnarray}}

\def\fnl{f_{\rm NL}}
\def\fnlloc{f_{\rm NL}^{\rm Loc}}
\def\fnleq{f_{\rm NL}^{\rm Eq}}
\def\fnlD{f_{\rm NL}^{(\Delta)}}

\begin{document}

\pagenumbering{roman}
\begin{titlepage}
\baselineskip=15.5pt \thispagestyle{empty}

\bigskip\

\vspace{1cm}
\begin{center}
{\fontsize{20.74}{24}\selectfont \sffamily \bfseries Biasing and the search for primordial non-Gaussianity beyond the local type}
\end{center}

\vspace{0.2cm}
\begin{center}
{\fontsize{12}{30}\selectfont J\'er\^ome Gleyzes$^{\blacktriangledown,\bigstar}$, Roland de Putter$^{\blacktriangledown}$, Daniel Green$^{\blacklozenge}$,  Olivier Dor\'e$^{\blacktriangledown,\bigstar}$} 
\end{center}

\begin{center}
\vskip 8pt
\textsl{$^\blacktriangledown$ California Institute of Technology, Pasadena, CA 91125}
\vskip 7pt
\textsl{$^\bigstar$ Jet Propulsion Laboratory, California Institute of Technology, Pasadena, CA 91109}
\vskip 7pt
\textsl{$^\blacklozenge$ Department of Physics, University of California, Berkeley, CA 94720, USA}

\end{center}

\vspace{1.2cm}
\hrule \vspace{0.3cm}
\noindent {\sffamily \bfseries Abstract}\\[0.1cm]
Primordial non-Gaussianity encodes valuable information about the physics of inflation, including the spectrum of particles and interactions.  Significant improvements in our understanding of non-Gaussanity beyond Planck require information from large-scale structure.  The most promising approach to utilize this information comes from the scale-dependent bias of halos.  For local non-Gaussanity, the improvements available are well studied but the potential for non-Gaussianity beyond the local type, including equilateral and quasi-single field inflation, is much less well understood.  In this paper, we forecast the capabilities of large-scale structure surveys to detect general non-Gaussianity through galaxy/halo power spectra.  We study how non-Gaussanity can be distinguished from a general biasing model and where the information is encoded.  For quasi-single field inflation, significant improvements over Planck are possible in some regions of parameter space. We also show that the multi-tracer technique can significantly improve the sensitivity for all non-Gaussianity types, providing up to an order of magnitude improvement for equilateral non-Gaussianity over the single-tracer measurement.
\vskip10pt
\hrule
\vskip10pt

\vspace{0.6cm}
\end{titlepage}

\thispagestyle{empty}
\setcounter{page}{2}
\tableofcontents

\clearpage
\pagenumbering{arabic}
\setcounter{page}{1}

%%%%%%%%%%%%%%
\section{Introduction}
%%%%%%%%%%%%%%

Understanding the physics of the very early universe is one of the central goals of modern cosmology.  Of particular interest is physics responsible for the creation of the initial fluctuations that seeded structure formation.  Primordial non-Gaussianity (PNG) is a powerful probe as it is sensitive to non-linear evolution of the fluctuations back to the time when the modes were created.  Different models of the early universe can produce qualitatively different predictions for the deviations from Gaussianty, meaning a detection or new upper limits will help determine the mechanism responsible.  Constraints on primordial non-Gaussianity from the CMB \cite{Ade:2015ava} currently provide the best limits on all types of non-Gaussanity, but fall short of some well-motivated targets.  Significant improvements are possible with large-scale structure (LSS) and can potentially reach some of these thresholds~\cite{Alvarez:2014vva}.

Scale-dependent halo bias is a powerful probe of primordial non-Gaussianity \cite{Dalal:2007cu,matver08,slosaretal08,desjseljak10}. In particular, scale-dependent bias is sensitive to the squeezed-limit bispectrum of primordial fluctuations. The single-field consistency relations \cite{Maldacena:2002vr,Creminelli:2004yq} state that in any single-field model of inflation, the dependence of this bispectrum on the wave vector of the long mode must be of the form (see also \cite{Kehagias:2013yd,Peloso:2013zw,Creminelli:2013mca,Creminelli:2013poa,Creminelli:2013nua} for more detailed discussions on consistency relations in the large-scale structure)
\beq
\langle\varphi(\k_L)\varphi(\q_1)\varphi(\q_2) \rangle' \propto k_L^2 P_{\rm p}(k_L)P_{\rm p}(|\q_1+\q_2|/2) \dots + \mathcal{O}(k_L^4),
\eeq
where $\varphi$ is the primordial Newtonian potential with power spectrum $P_{\rm p}$.  This primordial potential is related in linear theory to the Newtonian potential during matter domination by $\Phi(q) = D(z) T(k) \varphi(q)$ where $D(z)$ is the linear growth factor of matter fluctuations, normalized such that $D(z) = (1 + z)^{-1}$ during matter domination and $T(q)$ is the transfer function, which goes to unity as $q\rightarrow 0$.  The important feature is that the small-scale power is modulated by the primordial long-wavelength Newtonian potential.  As a result, any scale-dependent bias $\bfnl(k)$ for the halo density field due to primordial non-Gaussianity,  $\delta_h \supset \bfnl(k) \delta$, will be inversely proportional to the transfer function,
$\bfnl(k) \propto 1/T(k)$.  This is realized for equilateral type non-Gaussianity, since (see \cite{Schmidt:2010gw})
\beq
\bfnl(q) = 9 (b_\delta-1)\fnleq \Omega_m \delta_c \frac{H_0^2 R_*^2}{D(z)T(q) }\,,
\eeq
where $b_\delta$ is the linear bias, $\Omega_m$ is the density parameter of matter, $\delta_c$ is a critical density that typically appears in peak-background split calculations (see for example \cite{biasreview16}), $H_0$ is the Hubble parameter today and $R_*$ is the Lagrangian radius of the objects considered.
On the other hand, this consistency relation can be violated if multiple fields contributed to inflation \cite{Bartolo:2001cw,Bartolo:2004if,Lyth:2005fi,Vernizzi:2006ve}.
For example, local type non-Gaussianity, which arises when there are multiple light fields, gives
\beq
\bfnl(q) =3 (b_\delta-1)\fnlloc \Omega_m \delta_c \frac{H_0^2}{D(z)T(q)q^2}
\eeq
(in the single-field case, the consistency relation implies that any bias contribution with this $\propto q^{-2}$ scale dependence has to be exactly zero modulo projection effects \cite{dePutter:2015vga,Dai:2015jaa}).
More generally, models where one of the fields is not light also violate the consistency relation (see for example quasi-single field inflation \cite{Chen:2009zp}), and give
\beq
\bfnl(q) \propto \fnl^{\rm \Delta} \Omega_m \delta_c \frac{H_0^2}{D(z)T(q)q^2} (q R_*)^\Delta
\eeq
with $\Delta = 0 - 3/2$.

Constraints from scale-dependent halo bias on these types of primordial non-Gaussianity have been extensively studied in the literature, especially in the case of local non-Gaussianity (e.g. \cite{Leistedt:2014zqa,Dore:2014cca}), but also for equilateral \cite{baldaufetal16, giannanetal12,Byun:2014cea} and quasi-single field non-Gaussianity \cite{Norena:2012yi,Sefusatti:2012ye,Agarwal:2013qta}. One potential concern is that the scale-dependent bias effect may in principle be degenerate with contributions to halo bias from other sources than primordial non-Gaussianity, specifically non-linear bias and/or non-local\footnote{By this we mean that $\delta_h$ can depend on spatial derivatives of $\delta$.} bias (evidence for both of those has been in observed in simulations \cite{Chan:2012jj,Baldauf:2012hs}). While based on fundamental principles it is hard to mimic the $k_L^{-2}$ scale-dependence of local-type scale-dependent bias by such contributions, it may be an important effect in the other cases mentioned above. This can potentially severely weaken the constraining power of halo clustering on primordial non-Gaussianity. In particular, for equilateral non-Gaussianity, the scale-dependent bias can be expanded as $\bfnl(k) \propto 1/T(k) = 1 + (k/k_{\rm eq})^2 + \dots$, which at first sight is fully degenerate with the gradient bias expansion $b_\delta + b_{k^2} \, (R_* k)^2 + b_{k^4} \, (R_* k)^4 + \dots$. In fact, considering only the $k^2$ terms, and assuming $b_{k^2}$ to be of order unity, suggests an error floor on $\fnleq$ of order 1000 \cite{assassietal15,baldaufetal16}. On the other hand, the effects of equilateral non-Gaussianity and non-local bias should not be exactly the same, as they come in with a different characteristic scale. For primordial non-Gaussianity, the typical scale is $k_{\rm eq}$ ($\sim 1.6\times10^{-2} h/{\rm Mpc}$), while it is the Lagrangian size  of the halos of interest for the gradient bias ($R_*^{-1} \gtrsim 10^{-1} h/{\rm Mpc}$). In other words, one expects the degeneracy to not be exact when taking into account higher order terms.

In this paper, we study in more detail the constraints that can be obtained on primordial non-Gaussianity from scale-dependent bias in the presence of non-linear biasing and non-local (gradient) bias. We follow the approach of \cite{assassietal15,mcdonaldroy09,Baumann:2012bc} to write the most general bias expansion based on principles of symmetry and the equivalence principle. We then forecast constraints on primordial non-Gaussianity, marginalizing over the effects of the various bias contributions, where for non-linearities, we include all terms up to 1-loop order.

Of particular interest is how information about equilateral and quasi-single field-type non-Gaussianity is encoded in galaxy power spectra.  We study dependence on survey volume, number density, use of multiple tracers (as proposed in \cite{Seljak:2008xr}) to understand the reach and limitations of scale-dependent bias for non-local non-Gaussianity.  After marginalizing over bias, only quasi-single field inflation with $\Delta < 0.3$ and local non-Gaussianity show the potential for significant improvements over Planck \cite{Ade:2015ava} with realistic configurations.  Nevertheless, we show that the multi-tracer technique allows for significant improvements over single tracers and could extend the potential reach of this large-scale structure and scale-dependent bias.

The paper is outlined as follows:  In Section~\ref{sec:halo}, we detail the model for the halo power spectra in terms of a biasing model.    In Section~\ref{sec:method}, we explain our forecasting methodology.  In Section~\ref{sec:equil}, we focus on equilateral non-Gaussianity, because there one might expect degeneracy with other bias terms to be the strongest. We then turn to non-Gaussianity of the types that violate the consistency relations in Section~\ref{sec:beyond} (i.e.~require multiple fields in the early universe) and quantify the degradation to parameter constraints due to non-linear and non-local biasing. In Section~\ref{multi-tracer}, we discuss the multi-tracer technique and demonstrate the potential for improving constraints on equilateral non-Gaussianity.  We conclude in Section~\ref{sec:conc}.

The reader that is familiar with non-local, non-linear bias and/or MCMC forecasting may want to go directly to the results, starting in Section \ref{sec:equil}.

\section{Modeling the halo power spectrum}\label{sec:halo}

In this section, we describe the model of halo clustering. Our discussion here follows the systematic, general approach of \cite{Matsubara:2011ck,Matsubara:2012nc,assassietal15,Assassi:2014fva,mcdonaldroy09, schmidthetal13, mirabayietal15, sena15} (see also the review \cite{biasreview16} and references therein). We will consider the halo power spectrum in configuration space, i.e.~we will not include redshift space distortions, which would make the power spectrum anisotropic, and would significantky complicate the analysis.

\subsection{Scale-dependent bias from primordial non-Gaussianity}
\label{subsec:db}

We start with the leading order, local contributions to the halo overdensity $\delta_h$, including the scale-dependent bias from primordial non-Gaussianity, which is our signal,
\beq
\delta_h = b_\delta \, \delta + \fnl b_\Psi \, \Psi + \epsilon_0.
\eeq
Here, $\delta$ is the matter overdensity and $b_\delta$ the Gaussian, linear bias.
For non-Gaussian initial conditions, there can be a modulation of the small-scale variance of initial perturbations by a long-wavelength fluctuation,
\beq
\Psi(\q) \equiv \left(\frac{q}{\mu}\right)^\Delta \, \varphi(\q),
\eeq
where $\varphi$ is the primordial metric perturbation, related to the primordial curvature perturbation, and $\mu$ an arbitrary energy scale. The term $b_\Psi \, \Psi$ above describes the simplest form of bias due to this primordial non-Gaussianity (see \cite{assassietal15,assassietal15a} for additional refinements)
and has been tested in simulations \cite{wagnerverde12,scoccietal12,sefusetal12}. Finally, $\epsilon_0$ is a stochastic white noise contribution, which is assumed to be uncorrelated with the other terms.

The primordial metric fluctuation can be related to the present-day matter fluctuation by,
\beq
\delta(\q) = \mathcal{M}(q) \, \varphi(\q), \quad \text{with} \quad \mathcal{M}(q) \equiv \frac{2 q^2 T(q) D(z)}{3 \Omega_m H_0^2}.
\eeq
We can then write the non-Gaussian contribution as a scale-dependent bias,
\beq
\delta_h(\q) = \left( b_\delta + \bfnl(q) \right) \, \delta(\q) + \epsilon_0, \quad \bfnl(q) = b_\Psi \, \mathcal{M}^{-1}(q) \, \left(\frac{q}{\mu}\right)^\Delta.
\eeq
Parametrizing the primordial non-Gaussianity (specifically, the squeezed-limit bispectrum) in the conventional way, and applying a simple halo model, the scale-dependent bias can be written in terms of a non-Gaussianity parameter\footnote{Interestingly, these $\fnl$ parameters in a given volume may be biased relative to the true statistical $\fnl$~\cite{Baytas:2015nja} but the parameter $\Delta$ is robust to such effects~\cite{Bonga:2015urq}.} \cite{matver08,schmidtkam10,desjetal11a,desjetal11b,giannanetal12},
\bea
\bfnl(q) &=& 2 \, \fnlloc \, (b_\delta - 1) \, \delta_c \, \mathcal{M}^{-1}(q) \quad \text{(local)}\label{normLocal},  \\
\bfnl(q) &=& 6 \, \fnleq \, (b_\delta - 1) \, \delta_c \, (q \, R_*)^2 \, \mathcal{M}^{-1}(q) \quad \text{(equilateral)} \label{eq:db eq}, \\
\bfnl(q) &=& 6 \, \fnl^{(\Delta)} \, (b_\delta - 1) \, \delta_c \, (q \, R_*)^\Delta \, \mathcal{M}^{-1}(q) \quad \text{(generic exponent}\, \Delta \in[0,2]), \label{normEquiDelta}
\eea
where we recall that $\delta_c\,(=1.686)$ is the critical overdensity for spherical collapse, and $R_*$ is the Lagrangian radius of the halos of interest.\footnote{For the quasi-single field case, the conventional prefactor is not 6, but a more complicated function of $\Delta \in [0,3/2]$, see \cite{sefusetal12a}. When comparing our results to the existing QSF constraints in Section \ref{Sec: QSF constraints}, we will use the full normalization. However, when drawing a link from local to equilateral configurations, we will use the normalization \eqref{normEquiDelta}.}  The amplitudes of non-Gaussanity, $\fnl$, defined in eqs.~\eqref{normLocal}--\eqref{normEquiDelta} are determined by the amplitudes of the squeezed limit of the bispectrum.  Constraints from the CMB often normalize $\fnl$ by the amplitude of the bispectrum in the equilateral configuration, which can be non-trivially related to the amplitude in the squeezed limit.  This can be important for comparing limits from scale-dependent bias with limits from Planck, as we will see in the case of quasi-single field inflation (Section \ref{subsec:QSF}).

The above expressions for scale-dependent bias are based on the squeezed-limit behavior of the primordial bispectrum. This means their range of validity is technically limited to
$q/k_S \sim q \, R_* \ll 1$, where $q$ is the long mode, and $k_S$ the wave vector of the short modes.
For a given halo type, this places a requirement on the maximum wave vector included in the analysis, $k_{\rm max}$. In order to maximize the information from scale-dependent bias, we will push to relatively large values, $k_{\rm max} \, R_* \lesssim 1$, but one has to keep in mind that for modes that do not satisfy $q \, R_* \ll 1$, the actual form of the scale-dependent bias may be modified. We will come back to the optimization of $k_{\rm max}$ in Section \ref{subsec:gradientbias}.

\subsection{Non-linear and non-local halo bias}
\label{subsec:nonlinnonloc}

In this section, we will explicitly go step by step through the calculations of the halo-halo power spectrum in the presence of PNG. The reader not interested in those technical details is encouraged to skip to either the forecasting method Section \ref{sec:method}, or even directly to the results in Section \ref{sec:equil}.

Specifically, we want to compute
\be
\langle \delta_h(\k) \, \delta_h(\q) \rangle = (2 \pi)^3 \, \delta^{(D)}(\k + \q) \, P_{hh}(k),
\ee
using the methodology of standard perturbation theory (SPT). We refer the reader to \cite{Bernardeau:2001qr} for definitions and notations.
 
On top of the SPT contributions up to 1-loop, we will also include the most general bias model for $\delta_h$. To do so, we will allow $\delta_h$ to depend on any term that is permitted by the symmetries of the system, e.g.~rotational symmetry and the equivalence principle. This means that we allow $\delta_h$ to depend on derivatives of the matter overdensity $\delta$, leading to terms like $\nabla^2 \delta$, known as non-local bias. We will also include non-linear bias, i.e.~we will allow $\delta_h$ to depend on terms like $\delta^2$. Our derivation will follow closely the work of \cite{mcdonaldroy09} (and also \cite{Assassi:2014fva}), which computed $P_{hm} \propto \langle \delta(\k)\delta_h(\q) \rangle$ for gaussian initial conditions, to which we will add results of \cite{assassietal15} regarding the PNG.

In the expansion of $\delta_h$, the only term in $\Psi$ that we will keep is the linear one (i.e. the first line of eq (2.34) in \cite{assassietal15}), the other (loop) terms being much smaller for the scales of interest. Our expansion is going to be in $\delta_{\rm G}(\k)=\mathcal{M}(k)\varphi_{\rm G}(\k)$, where $\varphi_{\rm G}$ is a purely gaussian variable\footnote{In principle, the non-linear relation between the matter overdensity $\delta$ and $\delta_{\rm G}$ due to the PNG will show up as a modification to the kernels of standard perturbation theory. However we have checked that those modifications are negligible compared to the effect of the scale-dependent bias.}. Let's first discuss the structure of the terms that we will get. As in \cite{assassietal15}, we will conduct the expansion with diagrams for an easier representation.

\subsubsection{The ingredients}

The goal of this section is to present what are the terms that appear in the bias when considering that it is in principle non-local and non-linear. We will go through order by order (in $\delta_G$).

\begin{itemize}

\item \textbf{Linear in $\delta_{\rm G}$} 

Those terms will be represented by a line. External lines should be weighted by a linear $\delta_h$, given by 
\be
\label{eq:dhTree}
\delta_{ h}(q)=\left[b_\delta+\fnl \frac{b_\Psi}{\mathcal{M}(q)}\left(\frac{q}{\mu}\right)^\Delta+b_{q^2}q^2R_*^2+b_{q^4}q^4R_*^4+\ldots\right]\delta(\vec{q})\,.
\ee
The $b_{q^{2n}}q^{2n}R_*^{2n}$ terms come from Fourier transforming the dependence of $\delta_H$ on $(\nabla^{2})^{(n)}\delta$. The dots signify that in principle there are other terms in this gradient expansion. Internal lines on the other hand are necessarily $\delta_{\rm G}$. Contracting two lines gives a power spectrum $P_{\rm G}$ (represented by a black dot), with the appropriate bias weights. There is also a stochastic term, $\epsilon_0$, that is allowed by the symmetries. It does not correlate with $\delta_{\rm G}$ and gives a noise contribution to the power spectrum.

\item \textbf{Quadratic in $\delta_{\rm G}$} 

These will be represented by a striped circle with one line going in and two going out. The circle will have a weighting function $G(\k-\p,\p)$ (where $\k$ is the momentum of the line going in) that will depend on the exact nature of the quadratic term. Since our operators are expressed in terms of $\delta$ and not $\delta_h$, internal lines can only carry the standard perturbation theory kernel of second order for $\delta$, $F_{2}$. The restriction to the kernel $F_2$ will be shown by a circle with a grid. If the line is an external line, then two additional types of terms are possible: the quadratic overdensity $\delta^2$ and the square of tidal tensor $s_{ij}^2=\left(\nabla_i\nabla_j\Phi_\delta\right)^2-\frac{\delta^2}3$, where $\Phi_\delta\equiv \nabla^{-2}\delta$ .
Similarly to the linear case, there is a stochastic term $\epsilon_{\delta} \delta$ that appears. We will see that this term can be absorbed in the definition of the linear bias (see App.~\ref{detailsPS}).

\item \textbf{Cubic in $\delta_{\rm G}$} 

These will be represented by a striped square with one line going in and three going out. The square will have a weighting function $R(\q-\p_1-\p_2,\p_1,\p_2)$ (where $\q$ is the momentum of the line going in) that will depend on the exact nature of the cubic term. At one loop, on top of the standard perturbation theory kernel of third order for $\delta$, $F_3$, we have four additional contributions: $\delta^3$,  $ \delta\, s_{ij}^2$, $\nabla_i\nabla_j\Phi_\delta\nabla_j\nabla_k\Phi_\delta\nabla_k\nabla_i\Phi_\delta$. The last one, denoted $\text{Tr}\left[\Pi^{(1)}\Pi^{(2)}\right]$ in \cite{assassietal15} (it is related to the variable $st$ in \cite{mcdonaldroy09} although it is not exactly the same). This term is more subtle. At first order, $\delta$ and the velocity divergence $\theta$ are the same up to a normalization factor, which means that only a single variable is needed to describe them.  At second order they are not the same, but their difference (called $\Pi^{(2)}_{ij}$ in \cite{assassietal15}) is not a new variable, i.e.~it can be expressed in terms of the expressions in the paragraph right above (see \cite{assassietal15}). Only when multiplied with a first order quantity such as $\Pi^{(1)}_{ij}=\nabla_i\nabla_j\Phi_\delta$ does it become a new, independent variable, which is why it only starts appearing at cubic order. We will give more explicit expressions when looking directly at the kernels. There are also mixed stochastic terms, $\epsilon_{\delta^2} \delta^2$ and $\epsilon_{s^2} s_{ij}^2$, which give divergent contributions and therefore will not appear in the final (renormalized) result.
\end{itemize}

Now that we have the ingredients, it is a matter of getting all the possible terms in the expansion of $P_{hh}$ up to 1-loop. That will be represented by diagrams, similarly to standard perturbation theory \cite{Bernardeau:2001qr}. The terms in $P_{hh}$  are obtained by contracting the ``outgoing" lines from two ingredients in the list above. The contractions, that we will denote with a black dot, come with a linear power spectrum $P_{\rm G}$ and imposes the sum of the momenta of the lines to be zero. We will illustrate this below. Note that in principle, we should be working with renormalized quantities \cite{Assassi:2014fva} from the start. Instead, we will work with the bare quantities, and keep only the non-divergent parts in our final result.

\subsubsection{Tree level}
Only the terms that are at most linear in $\delta_{\rm G}$ can appear here. Thus, we have to contract two external lines as in eq.~\eqref{eq:dhTree}, as well as two stochastic terms $\epsilon_0$. This gives

\beq
P_{hh}^{\rm Tree}(q)\equiv b_{\rm full}^2P_{\rm G}(q)+P_{\epsilon_0}\,,
\eeq
where we have defined
\be
\label{btree}
b_{\rm full}(q)\equiv b_\delta+\fnl\frac{ b_\Psi}{\mathcal{M}(q)}\left(\frac{q}{\mu}\right)^\Delta+b_{q^2}q^2R_*^2+b_{q^4}q^4R_*^4+\ldots
\ee 
and we denoted the variance of the $\epsilon_0$ by $P_{\epsilon_0}$.

\subsubsection{1-loop}

\begin{figure}[h!]
\centering
\includegraphics[width=0.33\textwidth]{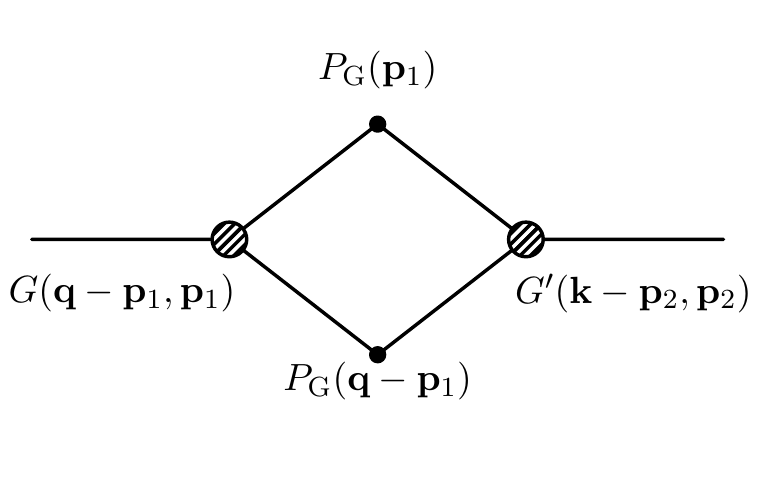}\includegraphics[width=0.33\textwidth]{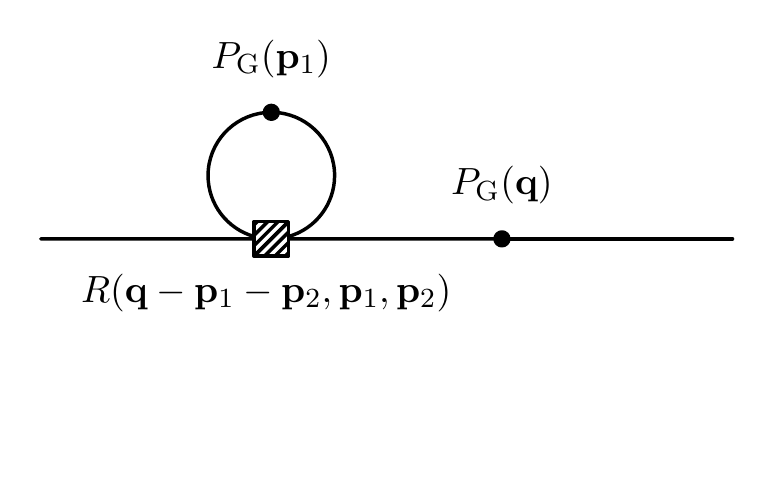}\includegraphics[width=0.33\textwidth]{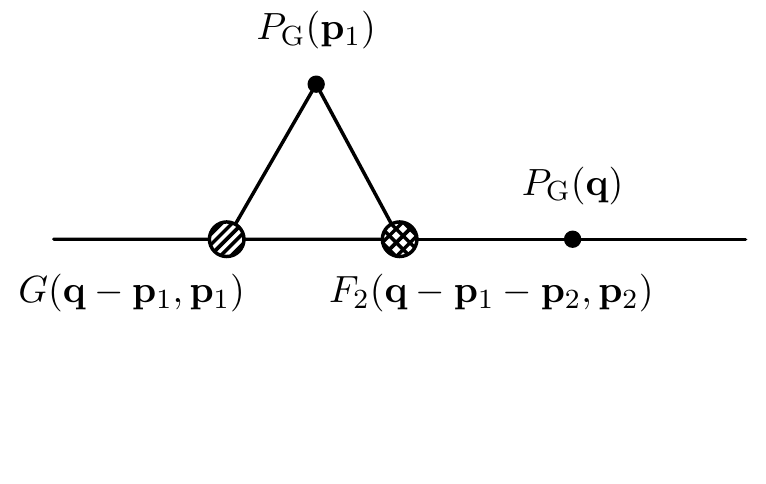}
\caption{\textit{Left}: Type 1) two second-order vertices. \textit{Middle}: Type 2) a cubic vertex and a linear $\delta_h$. \textit{Right}: Type 3) a second-order vertex and one SPT kernel $F_2$ contracted with $\delta_h$. The black dots represent the gaussian spectrum $P_{\rm G}$, and each external line carries a factor $b_{\rm full}$ (eq.~\eqref{btree}).}
\label{fig:Diagrams}
\end{figure}

Here, there are three types of diagrams that contribute. 1) a second-order vertex contracted with another second-order vertex (which is similar to the usual $P_{22}$ of SPT). 2)  a third order vertex with one loop and one contraction with a linear $\delta_h$ (the equivalent of $P_{13}$). There is a third type, that does not appear in SPT. This is because our quadratic operators are constructed from the full $\delta$, and not the first order $\delta_{\rm G}^{(1)}$ as in SPT. Therefore, we can have a second-order vertex contracted with the SPT second-order vertex $F_2$ and a linear $\delta_h$, which we call type 3. The structures of the diagrams are shown in Fig.~\ref{fig:Diagrams}

Let us now compute the contribution of the different types of diagrams.

\begin{itemize}
\item \textbf{Type 1)}

The contractions impose $\p_1=-\p_2$, so that there is only one integral over momentum (which is expected because we are looking at 1-loop). The diagram and its symmetrical partner therefore give

\beq
P_{hh}(q)\supset  \int_\p G(\q-\p,\p)\left[G'(\p-\q,-\p)+G'(-\p,\p-\q)\right]P_{\rm G}(|\q-\p|)P_{\rm G}(p)\,.
\eeq

We use the shorthand notation $\int_\p=\int {\rm d}^{3}p$. If $G=G'=F_2$, we recover $P_{22}$, the 1-loop term in SPT obtained when contracting two second-order $\delta_{\rm G}^{(2)}$.

\item \textbf{Type 2)}

The contractions impose $\p_2=-\q$. The diagram and its symmetrical partners (2 others, depending on which leg of the cubic vertex is contracted with the second external line) therefore give

\beq
P_{hh}(q)\supset P_{\rm G}(q) \int_\p P_{\rm G}(p)\left[R(-\p,\p,-\q)+R(-\p,-\q,\p)+R(-\q,-\p,\p)\right]\,.
\eeq

Again, if $R=F_3$, this is $P_{13}$, the 1-loop term in SPT obtained when contracting a first-order $\delta_{\rm G}^{(1)}$ with a third-order $\delta_{\rm G}^{(3)}$.

\item \textbf{Type 3)}

This type is slightly more involved. However, following the basic rules above, the diagram and its symmetrical partners (3 others, depending on where the second vertex is connected and where the second external line connects to the second vertex) give

\beq
P_{hh}(q)\supset 2P_{\rm G}(q) \int_\p P_{\rm G}(p) F_2(-\p,\q)\left[G(\q-\p,\p)+G(\p,\q-\p)\right]\,,
\eeq
where we have used that $ F_2(\k_1,\k_2)=F_2(-\k_1,-\k_2)=F_2(\k_2,\k_1)$.\footnote{In principle, there is another diagram where the external $\delta_h$ is connect to one branch of the first kernel $G$ while the two branches of the second kernel $F_2$ are connected together. However, this leads to a vertex $F_2(\p,-\p)$ which is zero.}

\end{itemize}

We therefore have the ingredients as well as the structure of the bias expansion. The only thing left is to assign the proper kernel to each ingredient, and then we will be able to compute $P_{hh}$ at 1-loop.

\subsubsection{The kernels}

Let us now translate the non-linear bias terms to quadratic $G(\k,\p)$ and cubic $R(\k,\p,\q)$ kernels. In order to compactly write the expressions, it is convenient to introduce the cosine between $\k_i$ and $\k_j$

\be
\mu_{ij}\equiv \frac{\k_i\cdot \k_j}{k_i k_j} \label{muij}\,,
\ee
as well as another cosine, between $\q-\p$ and $\p$
\be
\mu_{-}\equiv \frac{(\q-\p)\cdot \p}{|\q-\p| p}\,. \label{muminus}
\ee

\begin{itemize}

\item \textbf{Quadratic operators}: We have two kernels, $\delta^2$ and $s_{ij}^2=\left(\nabla_i\nabla_j\Phi_\delta\right)^2-\frac{\delta^2}3$. We do not consider $\delta \Psi$ and $\epsilon_\Psi \Psi$ as they are much smaller (suppressed by small non-Gaussianity and loop order).

We can therefore associate the kernels: $\delta^2\rightarrow 1$ and $s_{ij}^2\rightarrow G(\q-\p,\p)= \mu_{-}^2-1/3$, where $\mu_{-}$ is defined in eq.~\eqref{muminus}.

On top of it, there is the standard SPT kernel, $F_{2}$ (see App.~\ref{detailsPS} for details).

\item \textbf{Cubic operators}
We have the following kernels (they are symmetrized):

\beq
\begin{split}
\delta^3\rightarrow 1\,, \quad& \delta\, s_{ij}^2 \rightarrow R(\k_1,\k_2,\k_3)=\frac13\left(\mu_{12}^2+\mu_{23}^2+\mu_{13}^2-1\right)\,,\\
& \nabla_i\nabla_j\Phi_\delta\nabla_j\nabla_k\Phi_\delta\nabla_k\nabla_i\Phi_\delta\rightarrow \mu_{12}\mu_{23}\mu_{13}\,, \label{kernelcub}
\end{split}
\eeq
with $\mu_{ij}$ is defined in eq.~\eqref{muij}. The kernel for $\text{Tr}\left[\Pi^{(1)}\Pi^{(2)}\right]$  reads
\beq
\text{Tr}\left[\Pi^{(1)}\Pi^{(2)}\right]\rightarrow R( \k_1,\k_2,\k_3)=\frac13\left[\left(\frac{\k_1\cdot(\k_2+\k_3)}{k_1|\k_2+\k_3|}\right)^2\left[G_2(\k_2,\k_3)-F_2(\k_2,\k_3)\right]+{\rm perms}\right]\,,
\eeq
with $G_2$ the second-order kernel for the velocity (see \cite{Bernardeau:2001qr}). Plugging the expressions for $F_2$ and $G_2$, one finds

\beq
\text{Tr}\left[\Pi^{(1)}\Pi^{(2)}\right]\rightarrow R( \k_1,\k_2,\k_3)=\frac{2}{21}\left[\left(\frac{\k_1\cdot(\k_2+\k_3)}{k_1|\k_2+\k_3|}\right)^2\left[\mu_{23}^2-1\right]+{\rm perms}\right]\,.
\eeq

\end{itemize}

\subsubsection{The halo-halo power spectrum}

In this section, we will combine everything to compute $P_{hh}$ at first order in loops and $\fnl$. We have already computed the tree level contribution
\beq
\label{eq:Phhtree}
P_{hh}^{\rm Tree}(q)=b_{\rm full}^2P_{\rm G}(q)+P_{\epsilon_0}\,,
\eeq
with $b_{\rm full}$ defined in eq.~\eqref{btree}.

The next step is to remove the divergent quantities in the loop contributions (that would not have been there if we had worked with renormalized quantities from the beginning). Once this is taken into account (see App.~\ref{detailsPS}), the expression for the halo power spectrum reads

\beq
\label{PhhRen}
\begin{split}
P_{hh}(q)=\,&P_{hh}^{\rm Tree}(q)+b_{\rm full}(q)^2\,P^{\rm loop}(q)\\
+\,& 4 b_{\rm full}(q)\left\{\int_\p\left[ b_{\delta^2}+ b_{s^2}\left(\mu_{-}^2-\frac13\right)\right]\left[P_{\rm G}(|\q-\p|)-P_{\rm G}(p)\right]P_{\rm G}(p)\right\}\\
+\,&2 \int_\p\left[b_{\delta^2}+ b_{s^2}\left(\mu_{-}^2-\frac13\right)\right]^2\left[P_{\rm G}(|\q-\p|)-P_{\rm G}(p)\right]P_{\rm G}(p)\\
+\,&6b_{\rm full}(q)\, b_{\Pi \Pi^{(2)}}\,P_{\rm G}(q)\,\int_\p\left(\frac27\left[\left(\frac{\q\cdot\p}{q\,p}\right)^2-1\right]\frac{2\mu_{-}^2}3+\frac8{63}\right)P_{\rm G}(p)\,,
\end{split}
\eeq
where $P^{\rm loop}=P_{22}+P_{13}$ is the standard one loop contribution to the power spectrum \cite{Bernardeau:2001qr}. This expression is similar to the results in \cite{mcdonaldroy09}, with the addition of the PNG. Notice that out of the type 2) and 3) diagrams, only one remains after renormalization (last line). Indeed, when the divergent parts are removed, all the non-zero cubic terms are proportional to each other, as already noted in \cite{mcdonaldroy09}. This is why we regroup them into a single term, $b_{\Pi \Pi^{(2)}}$ (c.f.~Appendix \ref{detailsPS}). However, we cannot the same for the terms in second and third lines of eq.~\eqref{PhhRen} because $\mu_{-}$ depends on $\q$ and $\p$.

\begin{figure}[h!]
\centering
\includegraphics[width=0.65\textwidth]{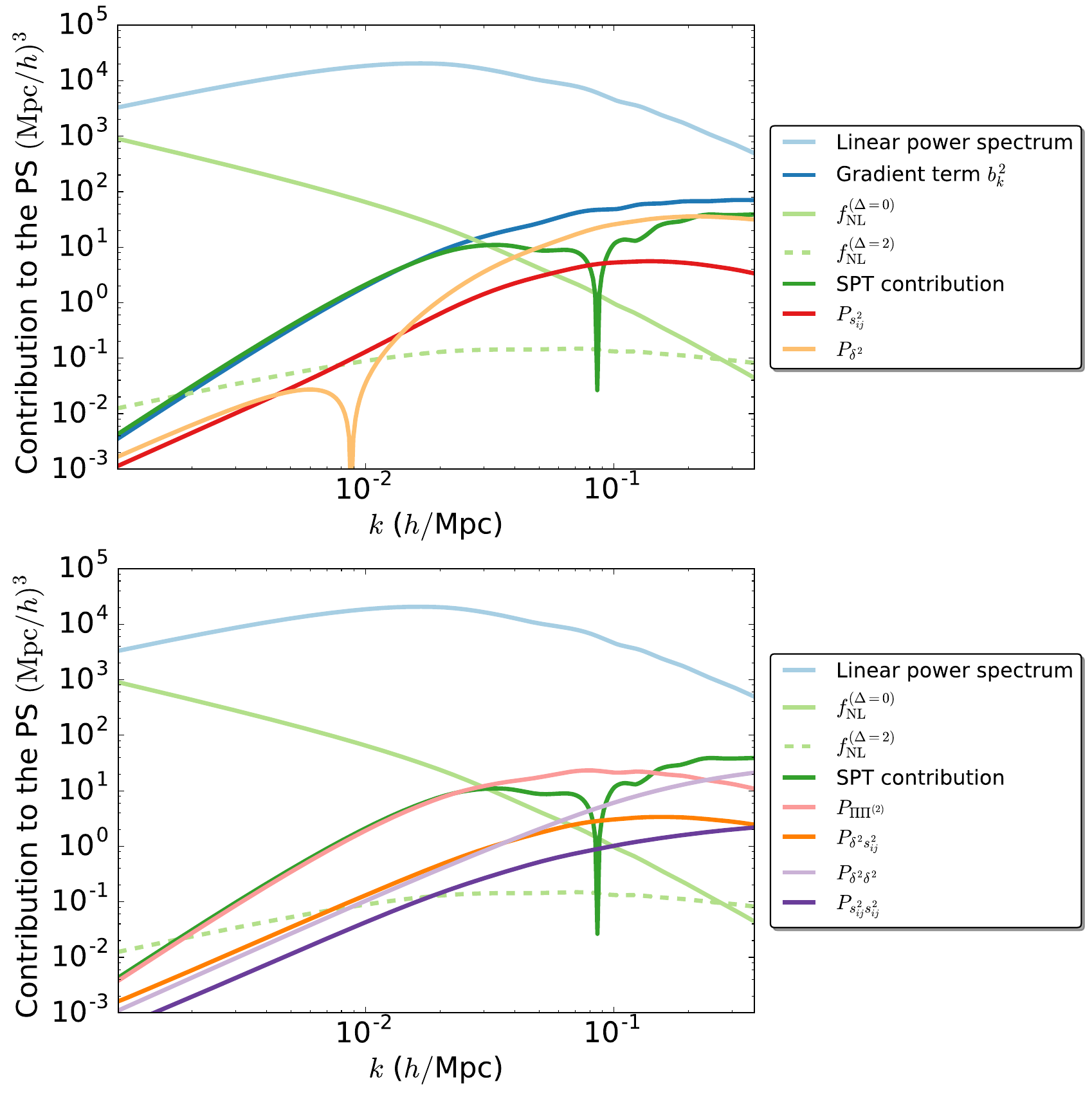}
\caption{The contributions to $P_{hh}$ compared to the linear power, $P_{\rm lin}=b_\delta^2 P_{\rm G}$.
%, using the one from CAMB \cite{Lewis:1999bs} for $P_{\rm G}$.
In the {\it top} panel, we show the terms linear in the extra bias coefficients, i.e.~$b_{k^2} k^2 R_*^2 P_{\rm G}(k)$. For the {\it bottom} panel, we also include non-linear contributions, such as $b_{\delta^2}^2P_{\delta^2\delta^2}$. There are also terms of the form $(b_{k^2} k^2 R_*^2)^2 P_{\rm G}(k)$, but we choose not to show them here for clarity. The linear bias is chosen to be $b_\delta=3.6$, the Lagrangian radius $R_*=3.7\, {\rm Mpc}/h$, the non-Gaussianity parameter $\fnl=1$ and all the other bias parameters are one. We show the scale-dependent bias contribution (linear in $f_{\rm NL}$ only) for $\Delta=0$ (local) and $\Delta=2$ (equilateral). At low $k$, the local ($\Delta=0$) contribution is the most important, and one can see by eye that no other bias terms can mimic its behavior. However, the $\Delta=2$ curve is both much smaller than $\Delta=0$ as well as more prone to degeneracies with non-local and non-linear contributions.}
\label{fig:pk}
\end{figure}

 To better gauge the size of those different terms, we plot them in Fig.~\ref{fig:pk} at $z = 1.5$, choosing $\fnl=1$, $b_\delta=3.6$, $R_*=3.7\,{\rm Mpc}/h$ and all the other bias parameters equal to one. The cosmology is set to that of Planck 2015 \cite{Ade:2015xua}, and the linear power spectrum is obtained from CAMB \cite{Lewis:1999bs}.

\section{Forecasting methodology: $\fnl$ constraints from the halo power spectrum}
\label{sec:method}

In order to estimate the constraining power of future surveys, we will resort to a Monte Carlo Markov Chain (MCMC), using the python package \project{emcee} \cite{ForemanMackey:2012ig}. To this end, we need the likelihood function, i.e.~the probability to observe a halo power spectrum $\hat{P}$ given a set of parameters $\btheta$. Since we want to focus on the bias expansion of eqs.~\eqref{btree} and \eqref{PhhRen}, we will fix all the cosmological parameters. Therefore, our set of parameters is given by
\be
\label{eq:thetaparam}
\btheta=\{b_\delta,b_{q^2},b_{q^4},\ldots,b_{s^2},b_{\delta^2},b_{\Pi\Pi},\Delta,P_{\epsilon_0}\}\,.
\ee

We approximate the likelihood by that of the power spectrum of a {\it Gaussian} galaxy density field, but using the full non-linear expression for the theory power spectrum, i.e.~the power spectrum $P_{\rm th}(k,{\bf \theta})$ given in eq.~\eqref{PhhRen}.
Given an observed (or in our case fiducial) spectrum $\hat{P}(k)$, this leads to the likelihood,
\be\label{eq:like}
\log{\mathcal{L}(\hat{P}(k)|\btheta)}=-\frac{N_k}2\left[\frac{\hat{P}(k)}{P_{\rm th}(k,\btheta)}-\log\left(\frac{\hat{P}(k)}{P_{\rm th}(k,\btheta)}\right)\right]\,,\quad N_k\equiv \frac{k^2 V}{2\pi^2}\Delta k\,,
\ee
where $N_k$ is the number of modes in a bin $[k,k+\Delta k]$ for a survey of volume $V$. We have ignored terms that do not vary with $\btheta$ and therefore will not modify our exploration of the parameter space.
The power spectrum thus is not a Gaussian variable, but follows a $\chi^2$ distribution.
This is particularly important for large scales, which are crucial for local non-Gaussianity,  as the number of modes is low so that the central limit theorem does not apply, cf.~\cite{Kalus:2015lna}.

\renewcommand{\arraystretch}{1.4}
\begin{table}[t]
\small
\begin{center}
\begin{adjustbox}{max width=\textwidth}
\begin{tabular}{|l|l||c|c|c|c|}
  \hline
Param. & Description & Fiducial & Prior \\
 \hline \hline
 $b_{\delta}$ &  Linear bias &  $  b(M,z) $ &$  [0,10] $  \\
 $b_{k^n}$ &  Gradient expansion &  $  0 $ &$  [-4,4] $  \\
 $b_{\delta^2}$ &  Quadratic bias &  $ 0 $ &$  [-4,4]  $  \\
 $b_{s^2}$ &  Tidal bias &  $  0 $ &$  [-4,4]  $  \\
 $b_{\Pi\Pi^(2)}$ &  Evolution bias &  $  0 $ &$  [-4,4]  $  \\
 $\fnl$ &  Non-Gaussianity &  $  0 $ &$  [-10^{5},10^{5}] $  \\
  $\Delta$ &  Type of PNG &  $  [0,2] $ &$  [0,2] $  \\
   $P_{\epsilon_0}$ &  Stochastic noise &  $  1/\bar{n}(M,z) $ &$  [0,2/\bar{n}] $  \\
 \hline
  \end{tabular}
\end{adjustbox}
\end{center}
\caption{ The parameters in $\btheta$ along with their fiducial values and priors. $b(M,z)$ and $\bar{n}(M,z)$ are respectively the bias and the mean density computed from the halo mass function at redshift $z$ for a minimum halo mass $M$. $b_{k^n}$ denotes generically all the terms in the bias expansions, whose truncation order will be specified later. For $\Delta$, the fiducial value will vary according the type of PNG that we want to constrain.}
\label{tb:MCMCsetup}
\end{table}

The cosmology is given by a $\Lambda$CDM model with parameters given by the latest Planck release \cite{Ade:2015xua}, and we will not vary them as we want to focus only on the potential degeneracy within the bias expansion. Our parameter estimations are for a survey of volume $V=100\, ({\rm Gpc}/h)^3$, with a mean redshift $z_{\rm m}=1.5$. We will consider modes going from $k_{\rm min}=2\pi/V^{1/3}$ to $k_{\rm max}=1/(2R_*)$\footnote{$k_{\rm max}$ is chosen so that the truncation of the gradient expansion $q R_*\ll1$ is sensible.}, with bin width $\Delta k=k_{\rm min}$. The fiducial linear bias $b_{\delta,\text{fid}}$, the lagrangian radius $R_*$ and galaxy density $\bar{n}$ (which gives the shot-noise and fixes the fiducial value of $P_{\epsilon_0}=1/\bar{n}$) are determined using the halo mass function and bias at redshift $z_{\rm m}$ \cite{tinkeretal08,tinkeretal10}. We will use a minimum halo mass of $M_{\rm h,min}=10^{13} M_\odot$ (see Section \ref{multi-tracer}, and in particular Fig.~\ref{fig:multi-tracer} for a discussion on this choice), which implies,
\be
b_{\delta,\text{fid}}=3.6\, \quad R_*=3.7\, \text{Mpc}/h\, \quad \bar{n}=1.2\times 10^{-4} (h/\text{Mpc})^3\,. \label{biasmhmin}
\ee
This gives a $k_{\rm max}=0.135\, h/\text{Mpc}$, well within the linear regime at redshift $z=1.5$.

We will take flat priors for all the free parameters. For the bias expansion, in particular the gradient terms, we will impose a flat prior between $[-4,4]$. Indeed, because we explicitly put the scaling with $R_*$ in the spatial derivatives, the terms in front of them should be of order one. This is why the expansion is not completely degenerate with $1/T(k)$ in the scale-dependent bias, which evolves on a different spatial scale given by the matter radiation equality, $k_{\rm eq}$. We will explore this in more details in the next section.

The fiducial parameter values and prior ranges are listed in Table \ref{tb:MCMCsetup}.
Although for realistic halos, the non-linear and non-local biases are expected to deviate from zero
(see \cite{lazeyrasetal16,mussoetal12,schmitettal15,Baldauf:2012hs} for assessment of a subset of these from simulations), we do not expect constraints on primordial non-Gaussianity to be very sensitive to the exact choice of fiducial values, so we set them to zero for simplicity.

The truncation order for the gradient expansion will be determined as the order when adding new gradient terms does not change the constraint on $\fnl$. As we will see, for $k_{\rm max}=1/(2R_*)$, this corresponds to including terms up to $k^{4}$.

\section{Equilateral non-Gaussianity}
\label{sec:equil}

We first consider constraints on scale-dependent bias due to equilateral non-Gaussianity, characterized by $\fnleq$.
According to eq.~(\ref{eq:db eq}), this type of non-Gaussianity leads to a scale-dependence
\beq
\bfnl(q) \propto \frac{1}{T(q)},
\eeq
where $T(q)$ is the tranfer function of matter perturbations, normalized to one when $q\rightarrow0$. Since the scale-independent part of $\bfnl(q)$ is unobservable (because completely degenerate with the linear bias), this means that the signal is dominated by small scales, $q > k_{\rm eq}$, where $k_{\rm eq} \approx 0.016 h/$Mpc is the matter-radiation equality scale. This stands in strong contrast with the more commonly studied halo bias due to local non-Gaussianity, which is dominated by large scales. This can readily be seen in Fig.~\ref{fig:pk}.

Since there is a large number of modes available on these small scales, in principle we expect scale-dependent bias to strongly constrain $\fnleq$. On the other hand, if we expand the scale-dependent bias signal around $q = 0$,
\beq
\label{eq:signaleq}
\bfnl(q) \propto 1 + c_2 \, \left( \frac{q}{k_{\rm eq}}\right)^2 + \mathcal{O}\left( \left( \frac{q}{k_{\rm eq}}\right)^4 \right),
\eeq
we find a scale-dependence that will to a certain extent be degenerate with the non-local and even non-linear bias terms discussed in Section \ref{subsec:nonlinnonloc}. However, the characteristic scale $k_{\rm eq}$ for the primordial non-Gaussianity signal is different that for the non-primordial terms $R_*^{-1}$ which are determined by the size of the halos. 
Thus, the degeneracy may not be exact, and below we investigate quantitatively what the expected constraints on $\fnleq$ are with and without the inclusion of non-linear and non-local biasing.

\subsection{Optimistic forecast, $b_\delta$ and $\fnl$ only}

\begin{figure}[h!]
\centering
\includegraphics[width=0.7\textwidth]{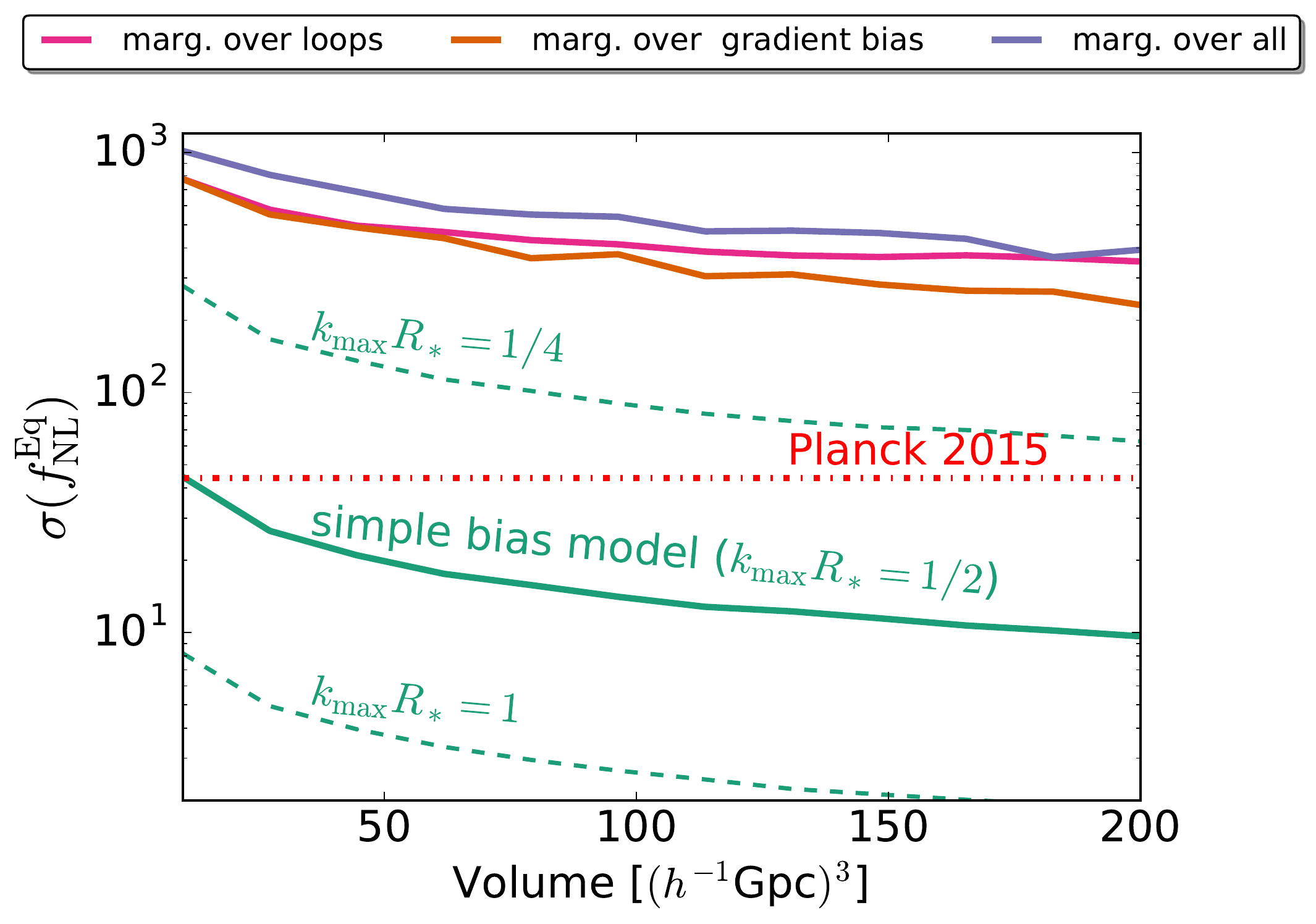}
\caption{Projected uncertainty on equilateral non-Gaussianity from scale-dependent bias as a function of survey volume. The solid curves show the results for our default survey assumptions and the default cutoff $k_{\rm max} \, R_* = 1/2$. The simple bias model (green) corresponds to only marginalizing over the linear bias parameter $b_\delta$. Applying a more general bias prescription leads to a severe degradation in constraining power, as seen with e.g.~the purple curve (all bias terms marginalized), which is about a factor of 40 larger. The dashed green curves explore the effect of varying the small-scale cutoff while keeping the simple bias model, and the red dotted-dashed line indicates the current constraint from the CMB bispectra.}
\label{fig:sigfnlvsvol}
\end{figure}

As a starting point, we consider constraints on $\fnleq$ marginalized only over the linear, Gaussian bias $b_\delta$. This corresponds  to the case of the other bias parameters being known to vanish. It is also the standard approach to forecasting constraints on local non-Gaussianity, which relies on very large scales, where additional bias corrections are small.
For the fiducial galaxy sample discussed in Section \ref{sec:method}, Figure \ref{fig:sigfnlvsvol} shows in green the uncertainty in $\fnleq$ as a function of the survey volume. The solid curves assume our default choice $k_{\rm max} \, R_* = 1/2$ and, for comparison, the dashed curves show the constraints for $k_{\rm max} \, R_* = 1/4, 1$ (we will discuss what is an appropriate cutoff choice in Section \ref{subsec:gradientbias}). Since there is a strong signal on small scales, the results are very sensitive to the choice of $k_{\rm max}$.

We also show in dashed red the current value of $\sigma(\fnleq)$ from the bispectra of CMB fluctuations, measured by Planck \cite{Ade:2015ava}. For our preferred choice $k_{\rm max} \, R_* = 1/2$, we see that a survey with volume $V \sim \mathcal{O}( (100 h^{-1} $Gpc$)^3 )$, comparable with next-generation galaxy surveys, appears to improve the $\fnleq$ constraint significantly compared to the CMB measurement: $\sigma(\fnleq)_{\rm LSS}\simeq 14$ vs $\sigma(\fnleq)_{\rm CMB}= 44$. Thus, scale-dependent bias in theory appears as a promising way of improving our knowledge of equilateral non-Gaussianity.

\subsection{Including non-local bias}
\label{subsec:gradientbias}

Next, we add gradient bias terms,
\beq
\label{eq:expandgradient}
\delta_h(\q) \supset \sum_{n \, {\rm even}}^{n_{\rm max}} b_{q^{n}} \, \left( R_* \, q\right)^{n} \, \delta^{(1)}(\q),
\eeq
see Section \ref{subsec:nonlinnonloc}, and marginalize over the bias parameters $b_{q^{n}}$. As discussed in the beginning of this section, these gradient bias terms come with a typically smaller characteristic length scale, $R_*$, than the characteristic scale for the onset of equilateral scale-dependent bias, $k_{\rm eq}^{-1}$. However, without restrictions on the coefficients in eq.~\eqref{eq:expandgradient}, if we simply expand both the signal, eq.~\eqref{eq:signaleq}, and the gradient bias in powers of $q^2$, it is clear that they are exactly the same, i.e.~we can mimic the effect of $\fnleq$ exactly with the gradient bias expansion by absorbing the difference in characteristic scales in the coefficients, $b_{q^n} \sim c_n \, (R \, k_{\rm eq})^{-n}$. This is a fine-tuning that does not reflect the physical difference between the two contributions. This physical difference is therefore imposed by our prior of order unity on the $b_{q^n}$ parameters, see Table \ref{tb:MCMCsetup}.

\begin{figure}[h!]
\centering
\includegraphics[width=0.65\textwidth]{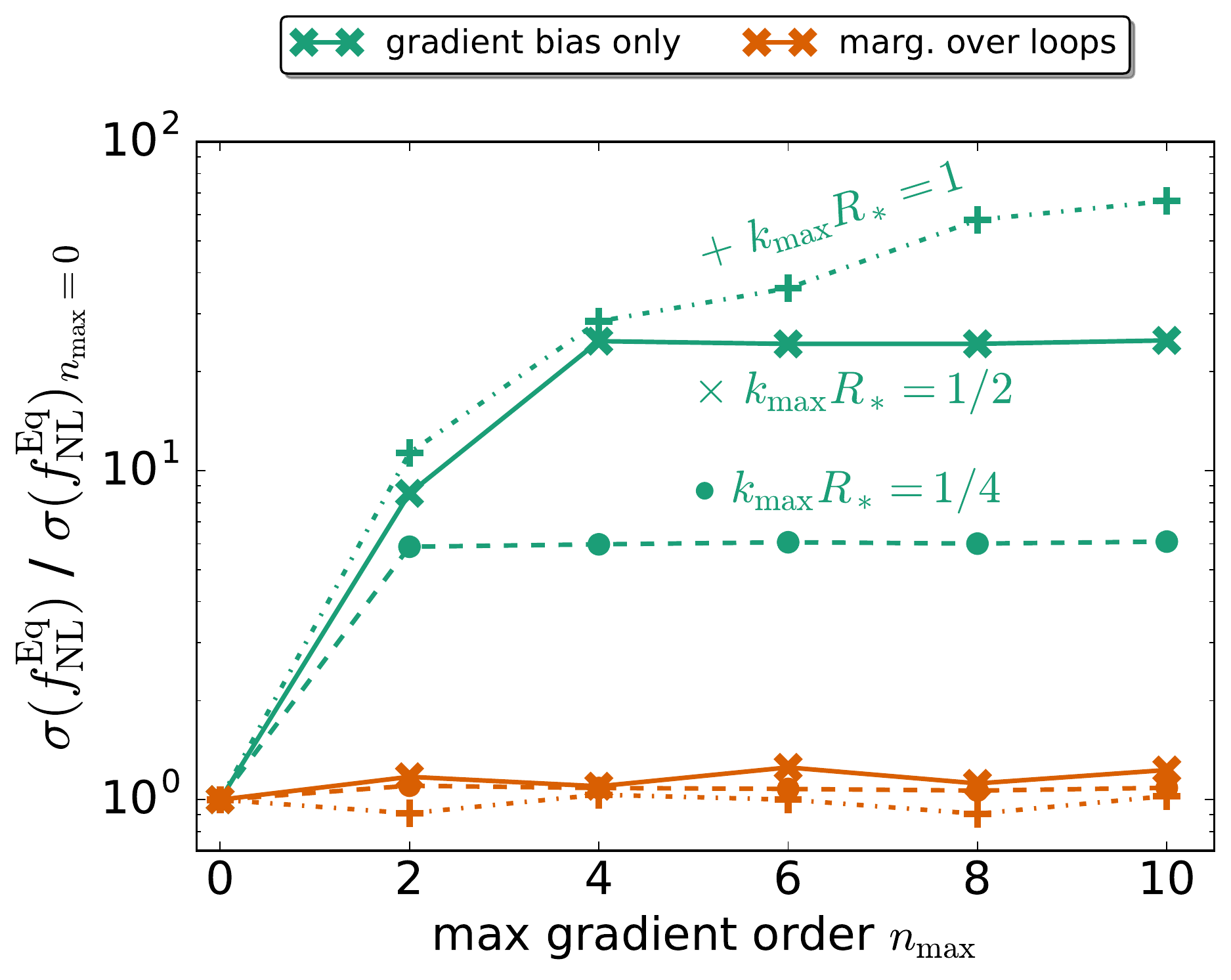}\;
\caption{Constraint on $\fnleq$ as a function of the order at which the gradient bias expansion is truncated, relative to the constraint in the absence of gradient bias (i.e.~with only the zeroth order term included). The green curves show the case without marginalizing over the loop terms in the tracer power spectrum, assuming short wavelength cutoffs $k_{\rm max} R_* = 1/4, 1/2, 1$ (circle, cross and plus markers respectively). For the default choice, $k_{\rm max} R_* = 1/2$, convergence is reached at $n_{\rm max} = 4$, which we will use as the truncation for most of this work. The orange curves show the same quantities, but including loop contributions. While the dependence on $n_{\rm max}$ is very different here, $n_{\rm max}=4$ is still a reasonable choice.}
\label{fig:sigfnlvsgradient}
\end{figure}

In general, the gradient bias expansion is expected to converge if the non-locality scale $\sim R_*$ associated with halo formation is small compared to the mode of interest, $q^{-1}$. Assuming this is satisfied, there must be some finite truncation of the expansion, $n_{\rm max}$, such that including terms beyond $n_{\rm max}$ negligibly affects the results, and in particular $\sigma(\fnleq)$. We study this convergence for $k_{\rm max} \, R_* = 1/4, 1/2, 1$ in Figure \ref{fig:sigfnlvsgradient}. The dots show the constraints using the model with $b_\delta, \fnleq$ {\it and} the gradient bias expansion as a function of $n_{\rm max}$, assuming a survey volume $V = 100 (h^{-1} $Gpc$)^3$.

We see that for $k_{\rm max} \, R_* = 1/2$, the results have converged by $n_{\rm max} = 4$. For $k_{\rm max} \, R_* = 1$, while the $\fnleq$ bound appears to converge by $n_{\rm max} = 8$, the cutoff $k_{\rm max}$ is clearly outside the regime where the gradient expansion can be expected to make sense. For smaller cutoff, $k_{\rm max} \, R_* = 1/4$, convergence is reached early, at $n_{\rm max} = 2$, but since this cutoff means using a much smaller number of modes, the $\fnleq$ constraint is significantly weakened. As a compromise between the gradient expansion being well behaved, and using as many modes as possible, we choose as our default value for the rest of this work the cutoff $k_{\rm max} \, R_* = 1/2$. As discussed in Section \ref{subsec:db}, our expressions for the scale-dependent bias assume the squeezed-limit bispectrum behavior, which technically is only valid for $q \, R_* \ll 1$. We thus note that for the smallest modes we include, there may be non-negligible corrections to the scale-dependent bias that we do not take into account.

Assuming the default cutoff $k_{\rm max}$ motivated above, the orange curve in Figure \ref{fig:sigfnlvsvol} shows the constraint on $\fnleq$ as a function of survey volume when including the gradient expansion. We see that marginalization over non-local bias severely weakens the bound on non-Gaussianity, by about a factor of $25$.

\subsection{Including non-linearities}

Finally, we include the loop terms from non-linear bias and evolution, discussed in Section \ref{subsec:nonlinnonloc}, marginalizing over the corresponding parameters. The orange crosses in Figure
\ref{fig:sigfnlvsgradient} confirm that even in this more general scenario, our truncation of the gradient expansion at $n_{\rm max} =  4$ is appropriate. The purple curve in Figure \ref{fig:sigfnlvsvol} shows the final constraints as a function of survey volume. The degradation factor relative to the simple case without non-linear or non-local bias is about $40$, slightly worse than with non-local biasing only. We note that it would be wrong to conclude from this that non-local bias is necessarily more degenerate with scale-dependent bias than the non-linear terms. If we add the non-linear terms first, and {\it then} include the non-local terms, we would again find that the first step gives the biggest deterioration (see the magenta curve in Fig.~\ref{fig:sigfnlvsvol}).

\subsection{Summary}

While a ``naive'' forecast suggests scale-dependent bias in galaxy surveys may improve the constraint on $\fnleq$ beyond the current CMB error bar to $\sigma(\fnleq)\simeq 20$, implementing a general treatment of bias and non-linear evolution leads to strong degeneracies, weakening the constraints by roughly a factor $40$ leading to $\sigma(\fnleq)\simeq 500$ for a survey volume $V = 100 (h^{-1} $Gpc$)^3$.
On the bright side, we do find that the degeneracy is not exact, and that even after marginalization, information on $\fnleq$ remains in scale-dependent bias.
Unfortunately, the degraded uncertainties are at least an order of magnitude above the current Planck constraint, even assuming very large survey volumes of order a few $\times 100 (h^{-1} $Gpc$)^3$.

The forecasts in this section assume the use of a single halo sample. When multiple samples with different biases and with high number density are available, it is possible to evade the cosmic variance bound and to strongly improve the constraints \cite{Seljak:2008xr,Yamauchi:2014ioa}. We will study in Section \ref{multi-tracer} if this method can make constraints on $\fnleq$ from scale-dependent bias more competitive with the CMB. We also point out that for equilateral non-Gaussianity the halo bispectrum \cite{Sefusatti:2007ih,Mizuno:2015qma,Hashimoto:2016lmh} is expected to do better than the halo power spectrum because it makes use of primordial bispectrum configurations beyond the squeezed limit, and the equilateral signal is dominated by those triangles (unlike local non-Gaussianity). Even for the halo bispectrum, however, degeneracies with non-linear evolution and biasing place limits on what can be achieved \cite{baldaufetal16}.

\section{Beyond single-field inflation} \label{sec:beyond}

\subsection{Particle physics and the squeezed limit}

Inflation is thought to have occurred at energies such that $H \lesssim 10^{14}$ GeV.  During inflation, any particles with masses $m \lesssim H$ are excited from the vacuum and can potentially impact the evolution of the fluctuations we ultimately observe.  These extra fields are particularly important in the squeezed limit as the can lead to violations of the single-field consistency relations~\cite{Chen:2009zp,Baumann:2011nk}.  There are good reasons to think a plethora of new particles could appear at these energies and we should take this possibility seriously~\cite{Chen:2009zp,Baumann:2011nk,Assassi:2013gxa,Craig:2014rta,Arkani-Hamed:2015bza,Lee:2016vti}.

The most well-studied possibility is local non-Gaussianity, which arises most commonly from scalar fields with $m \ll H$. In this case, the primordial perturbation of the gravitational potential $\varphi$ can be written as
\be
\varphi=\varphi_{\rm G}+\fnl(\varphi_{\rm G}^2-\langle\varphi_{\rm G}\rangle^2)\,,  
\ee
where $\varphi_{\rm G}$ is a gaussian variable of power spectrum $P_{\varphi}$.
In the squeezed limit, local non-Gaussianity takes the form
\beq
\lim_{k_1 \to 0} B(k_1,k_2,k_3) =\langle\varphi(k_1)\varphi(k_2)\varphi(k_3)\rangle\to 4 \fnl P_{\varphi}(k_1) P_{\varphi}(k_2) \,
\eeq
where $k_1 \to 0$ forces $k_2 \simeq k_3$.  For $m \simeq H$ (known as Quasi-Single Field, or QSF), the extra fields do not directly contribute to the dynamics at the end of inflation (often responsible for local non-Gaussianity) but lead to non-trivial mode coupling at horizon crossing.  This leads to a squeezed\footnote{This expression is only valid for $\Delta\neq 3/2$. The expression for $\Delta=3/2$ can be found in e.g. \cite{sefusetal12a}.}  bispectrum \cite{Chen:2009zp,sefusetal12a}
\beq
\label{squeezedBS}
\lim_{k_1 \to 0} B(k_1,k_2,k_3) \to- \frac{18\sqrt{3}}{\pi}\frac{\Gamma(3/2-\Delta)}{2^\Delta\, N_{3/2-\Delta}(8/27)}  \fnl^{(\Delta)} \left(\frac{k_1}{k_2} \right)^{\Delta} P_{\varphi}(k_1) P_{\varphi}(k_2) \,
\eeq
where $\Delta  = \frac{3}{2} - \sqrt{\frac94-\frac{m^2}{H^2}}$ and $N_{\nu}$ is the second-kind Bessel function of order $\nu$.  For $m \leq 3 H/ 2$, we find weaker power law scaling in the squeezed limit, with $0 <\Delta \leq \frac{3}{2}$.  For $m > 3 H/2$ this behavior becomes oscillatory and the amplitude is exponentially suppressed~\cite{Noumi:2012vr,Arkani-Hamed:2015bza}.  In both cases, the overall power law is determined by the time evolution of the wave-function for the massive fields outside the horizon.  The suppression in the squeezed limit is ultimately due to the decay of the massive field between the time of horizon crossing of the short and long wavelength modes~\cite{Baumann:2011nk}.  In these simple models, the power law never reaches the single-field limit $\Delta = 2$ because we must reproduce the $a^{-3}$ dilution expected for very massive particles~\cite{Arkani-Hamed:2015bza}.  However, by using interactions, we can modify the time evolution to achieve any $0 \leq \Delta \leq 2$~\cite{Green:2013rd}.  

\subsection{Constraints on $f_{\rm NL}^{(\Delta)}$ ($\Delta = 0 - 2$)}
\label{Sec: QSF constraints}

Here we study constraints on scale-dependent bias with arbitrary scale-dependence, $\Delta = 0 - 2$, where, in summary, $\Delta = 0$ corresponds to local non-Gaussianity, $\Delta = 2$ to equilateral non-Gaussianity, and the range $\Delta =  0 - \tfrac{3}{2}$ to quasi-single field inflation (QSFI) with a range of masses.

In order to understand the statistical power of scale-dependent bias to constrain these models, it is useful to consider the constraints where the amplitude in the squeezed limit is fixed as a function of $\Delta$.  Specifically, in this section we will assume
\bea
\bfnl(q) &=& 6 \, \fnl^{(\Delta)} \, (b_\delta - 1) \, \delta_c \, (q \, R_*)^\Delta \, \mathcal{M}^{-1}(q) \, \ ,    
\eea
where we will take $\fnl^{(\Delta)}$ to vary independently of $\Delta$.  The case $\Delta = 0$ corresponds to local non-Gaussianity, although with a slightly unusual normalization $3 \fnl^{(\Delta = 0)} =  \fnlloc$.  In this limit, the signal-to-noise is dominated by the largest scales where they are robust to non-linear evolution and biasing.  However, because of the non-trivial $q$-dependence of $\mathcal{M}^{-1}(q)$ there is also potentially information available at large $q$ that is not degenerate with the bias parameters.  As we increase $\Delta> 0$, the amount of information available at $q \to 0$ decreases and (naively) increases at $q \to k_{\rm max}$.

\begin{figure}[h!]
\centering
\includegraphics[width=0.65\textwidth]{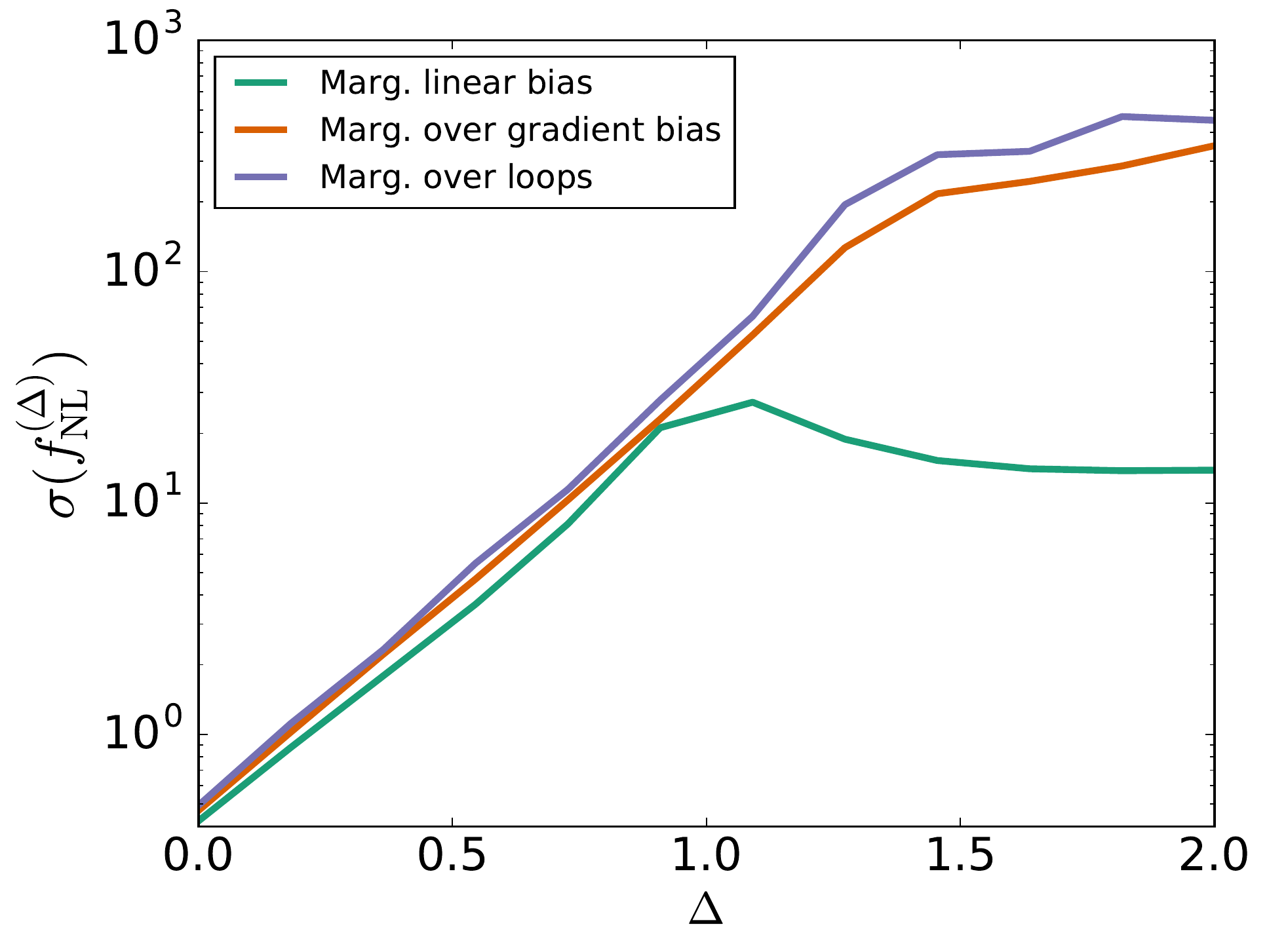}\
\caption{The evolution of $\sigma(\fnl^{(\Delta)})$ as a function of a fixed $\Delta$, for different marginalization schemes. The green curve is the naive scenario, where marginalizing solely over the linear bias $b_\delta$. For the orange curve, we also marginalized over gradient terms, but keeping the non-linear bias zero. The purple curve shows the full marginalization, over the every term in the expression \eqref{PhhRen}. Up to $\Delta=1$ the behavior of the three curves is similar, and the increase with $\Delta$ comes from the fact the scales that dominate the signal get away from the squeezed limit, which is what we probe best with scale-dependent bias. At $\Delta=1$, the degeneracies start to play a more important role, as the signal is dominated by the small scales which are more non-linear and non-local. This explains the large degradation from the green to the purple curve.}
\label{fig:DeltaVaryings}
\end{figure}

The constraint on $\fnl^{(\Delta)}$ for varying $\Delta \in [0,2]$ is shown in Figure~\ref{fig:DeltaVaryings}.  We see that when $\Delta < 1$, the results are unaffected by marginalization over bias parameters, which means most of the constraining power is coming from large scales $q \ll k_{\rm max}$ (for example, one can see in Fig.~\ref{fig:pk} that the other contributions to $P_{hh}$ are very different from that of $\fnl^{(\Delta=0)}$).  However, for $\Delta > 1$, we find much weaker constraints when marginalizing over higher order biases.

We can understand the forecasts qualitatively from the likelihood function in eq.~\eqref{eq:like}.  For a small deviation from the hypothetical model, $\hat P = P_{\rm th} + \delta P$ then we have
\be\label{eq:likeapprox}
\log{\mathcal{L}(\hat{P}(k)|\btheta)}\approx-\frac{k^3 V}{4\pi^2}\Delta \ln k \, \frac{1}{2} \left(\frac{ \delta P(k)}{P_{\rm th}(k,\btheta)}\right)^2 \ .
\ee
If we consider only the change from the primordial non-Gaussanity, we have $\delta P =  12 b_\delta(b_\delta - 1) \fnl^{(\Delta)}  \, \delta_c \, (k \, R_*)^\Delta \, \mathcal{M}^{-1}(k) P(k)$ and therefore the likelihood scales like
\beq
\log{\mathcal{L}(\hat{P}(k)|\btheta)}\approx - \frac{18 k^3 V}{\pi^2}\Delta \ln k \frac{(b_\delta-1)^2}{b_\delta^2} \fnlD{}^2 \delta_c^2 (k \, R_*)^{2 \Delta} \, \mathcal{M}^{-2}(k) \ ,
\eeq
where we have assumed a sample-variance limited measurement.
As $k \to 0$, $\mathcal{M} \propto k^2$ and we see therefore that for $\Delta \leq 1/2$, the signal-to-noise at low-$k$ is dominated by the smallest $k$ (largest scales).  However, $\mathcal{M}(k)$ is not simply a power law and for $k > k_{\rm eq}$ the $\mathcal{M}(k)$ grows more slowly than $k^{3/2}$ and therefore the signal-to-noise increases with increasing $k$.  As a result, we expect there to be contributions to the signal both at largest and smallest scales in the survey for $\Delta < 1/2$.  Increasing $\Delta$ makes the contribution to the signal increasingly dominated by the smallest scales, especially for $\Delta > 1/2$.

Thus, the best constraints are for $\Delta$ close to zero, where adding other bias terms does not significantly alter $\sigma(\fnl)$. This is because for low $\Delta$, the signal peaks at large scales which are linear and do not get large contributions from the rest of the bias expansion (see Fig.~\ref{fig:pk}).  By contrast, small scales are the most non-linear and degeneracies with biasing will appear as weakening constraints. Therefore, when approaching the equilateral case, the constraints notably worsen for the full bias expansion. The peak in the curve when marginalizing solely over the linear bias comes from the fact that for $k$ close to $k_{\rm max}$ (where most of the signal is for these values of $\Delta$), the transfer function $T(k)$ roughly behaves as $k^{-1}$, so that $\mathcal{M}(k)\propto k$. Thus, for $\Delta=1$, there is enhanced degeneracy with the linear bias.  As we increase $\Delta > 1$, it will become less degenerate with $b_\delta$ but more degenerate with the higher order biases, which is reflected in the deferences in the forecasts for difference biasing models.

The scaling properties for these forecasts follow from eq.~\eqref{eq:likeapprox} and, in particular, the noise level set by the cosmic variance of the tracer power spectrum.  However, mode-coupling occurs on a mode-by-mode basis and can, in principle, be measured without cosmic variance.  We will discuss the multi-tracer approach to cosmic variance cancellation in Section~\ref{multi-tracer}, but we should anticipate that the forecasts with behave qualitatively differently because of the change of the noise properties.

\subsection{Constraints on quasi-single field inflation}
\label{subsec:QSF}
Having understood the scaling behavior of the signal-to-noise for generic scale-dependent bias, we are now in a position to understand the corresponding constraints on QSFI.  We now must take into account the change to the normalization of the squeezed limit, given by eq.~\eqref{squeezedBS}.

Planck has put constraints on non-Gaussianity coming from QSF inflation \cite{Ade:2015ava}, using an expression for the bispectrum that interpolates between the local and equilateral shapes. The expression is given by \cite{Chen:2009zp}
\be
\label{eq:tempQsA}
\begin{split}
B_{\varphi}(k_1,k_2,k_3)&\equiv 6\, C_\varphi^2\, F(k_1,k_2,k_3)\,,\\
F(k_1,k_2,k_3)&\equiv\frac{3^{3/2}}{N_\nu(8/27)}\fnl\frac{N_\nu\left[8 k_1 k_2 k_3/(k_1+k_2+k_3)^3\right]}{(k_1 k_2 k_3)^{3/2}(k_1+k_2+k_3)^{3/2}}\,,
\end{split}
\ee
where  $\nu=3/2-\Delta$. When looking at the squeezed limit\footnote{large-scale structure probes of quasi-single field inflation need not use the squeezed limit to place constraints on $\fnl$. See~\cite{Shandera:2013mha,Dimastrogiovanni:2015pla, Meerburg:2016zdz} for some examples.  }, one recovers eq.~\eqref{squeezedBS}, which leads to a scale-dependent bias similar to eq.~\eqref{normEquiDelta}
\be
\label{QSFApp}
\lim_{k \to 0}\bfnl^{\rm QSF}(k)= -\frac{9\sqrt{3}}{\pi}\frac{\Gamma(3/2-\Delta)}{2^\Delta\, N_{3/2-\Delta}(8/27)} \, \fnl^{(\Delta)} \, (b_\delta - 1) \, \delta_c \, (q \, R_*)^\Delta \, \mathcal{M}^{-1}(q)\,,
\ee
which, as before, is only valid for $\Delta \neq 3/2$. It diverges as $\Delta\to 3/2$, because the limit $k\to 0$ does not commute with the limit $\Delta\to 3/2$. There is a different expression in this case, given in \cite{sefusetal12a}. This expression is mainly for comparison with eq.~\eqref{normEquiDelta}. For our analysis however, we will not use it. Indeed, we expect that for large values of $\Delta$, the signal peaks at scales where this approximation is not valid. This is why we will rather use the expression computed from integrating over the bispectrum \eqref{eq:tempQsA} without taking the squeezed limit. This means that it is given by (see \cite{sefusetal12a})
\beq\label{eq:dbsd}
\bfnl(k)= \frac{\delta_c\left[b_\delta-1\right]}{2 \mathcal{M}(k)}\frac{I_{21}(k)}{\sigma^2_m}\,,
\eeq
where
\begin{align}
\sigma^2_m&\equiv\int {\rm d}^3q\, P_{\rm G}(q)W_R(q)^2\,,\\
I_{21}(k)\equiv\frac{1}{P_\varphi(k)}\int {\rm d}^3q\, \mathcal{M}(q)&\mathcal{M}(|\q-\k|)W_R(q)W_R(|\q-\k|)B_\varphi(q,|\q-\k|,k) \label{eq:I21}\,,
\end{align}
and $W_R(q)=3\left[\sin(qR)-qR\cos(qR)\right]/(qR)^3$ is the Fourier transform of a top-hat filter.

We want to use this form of the scale-dependent bias to compare with Planck, which conducted an analysis of QSFI in \cite{Ade:2015ava}. Note that in their case, they used directly the bispectrum, not only the squeezed limit as we are doing (even with the bias given by eq.~\eqref{eq:dbsd}). Therefore, we do not expect to improve their constraints for large values of $\Delta$, where equilateral configurations receive a sizable part of the signal.

\begin{figure}[h!]
\centering
\includegraphics[width=0.65\textwidth]{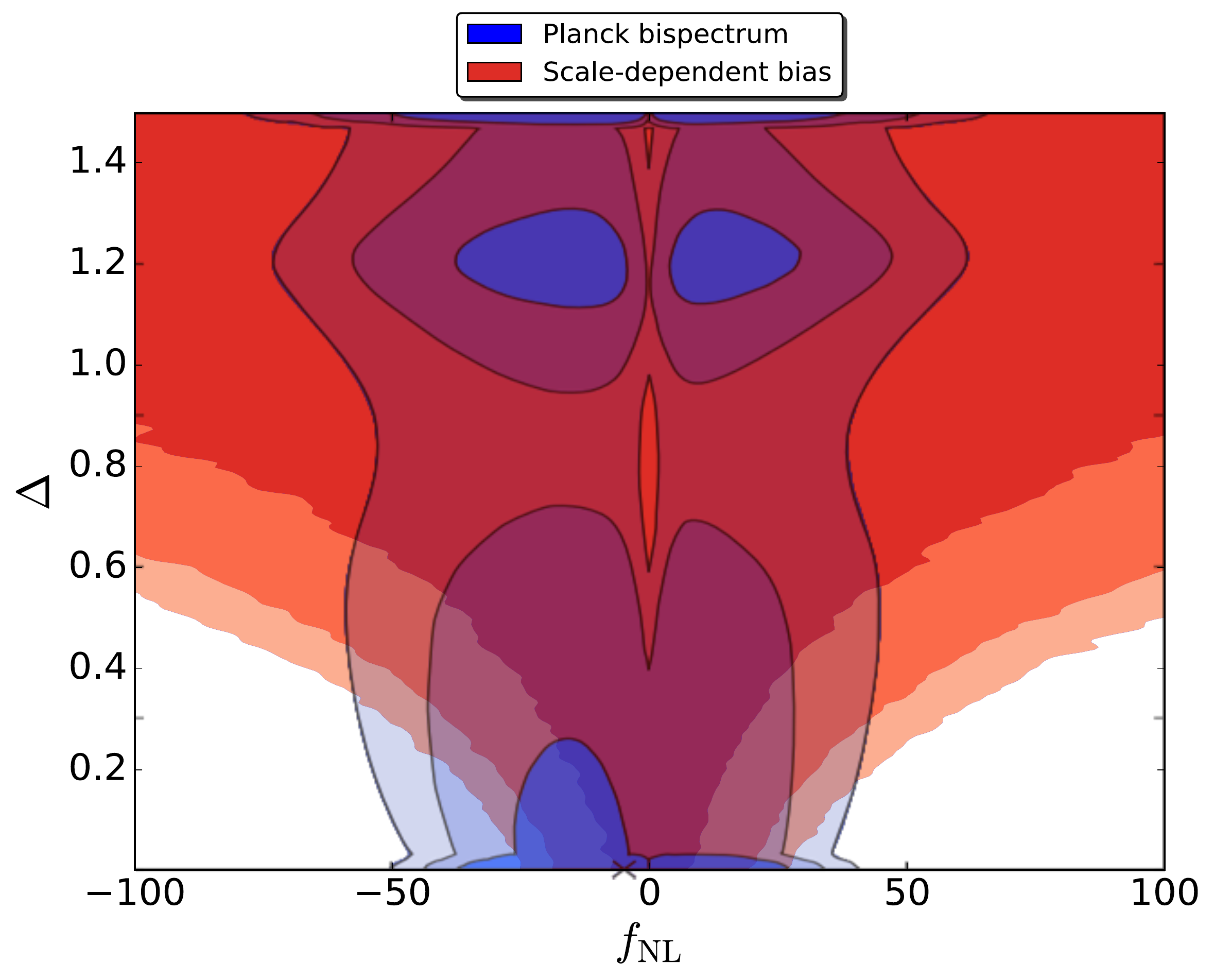}\, 
\caption{The red regions are the $68\%$, $95\%$ and $99\%$ contours from our MCMC runs. The blue are the $68\%$, $95\%$ and $99\%$ contours from Planck \cite{Ade:2015ava}. At low $\Delta$, below $\sim 0.3$, the scale-dependent bias does better than Planck. Indeed, in this case, the signal is dominated by the largest scales, where galaxy surveys have a decisive advantage. However, as $\Delta$ increase, our constraints keep on worsening because the signal is dominated by smaller scales, where the non-local and non-linear bias terms have significant contributions. These are also the scales at which one expects that the scale-dependent bias, that relies on squeezed configurations of the bispectrum, should perform poorly compared to the bispectrum. Indeed, for Planck, high $\Delta$ values just means that the signal is more dominated by equilateral shapes, which is well captured by the bispectrum. This is why the blue contours do not change significantly as a function of $\Delta$.}
\label{fig:QSFPlanck}
\end{figure}

We show the results in Fig.~\ref{fig:QSFPlanck}, where we have let $\Delta$ be a free parameter, with flat prior between $[0,3/2]$ and fiducial values $\Delta=0$ and $\fnl=0$.

As we saw in the previous section, as we increase $\Delta$ away from zero we slowly interpolate between being dominated by the largest scales to dominated by the smallest.  It is therefore not surprising that Fig.~\ref{fig:QSFPlanck} shows that we have better constraints than existing Planck limits on QSFI for $\Delta \lesssim 0.3$. However, as we increase $\Delta$, our constraints keep on weakening, while those of Planck stay roughly the same. This again was expected from Fig.~\ref{fig:DeltaVaryings}.

\begin{figure}[h!]
\centering
\includegraphics[width=0.65\textwidth]{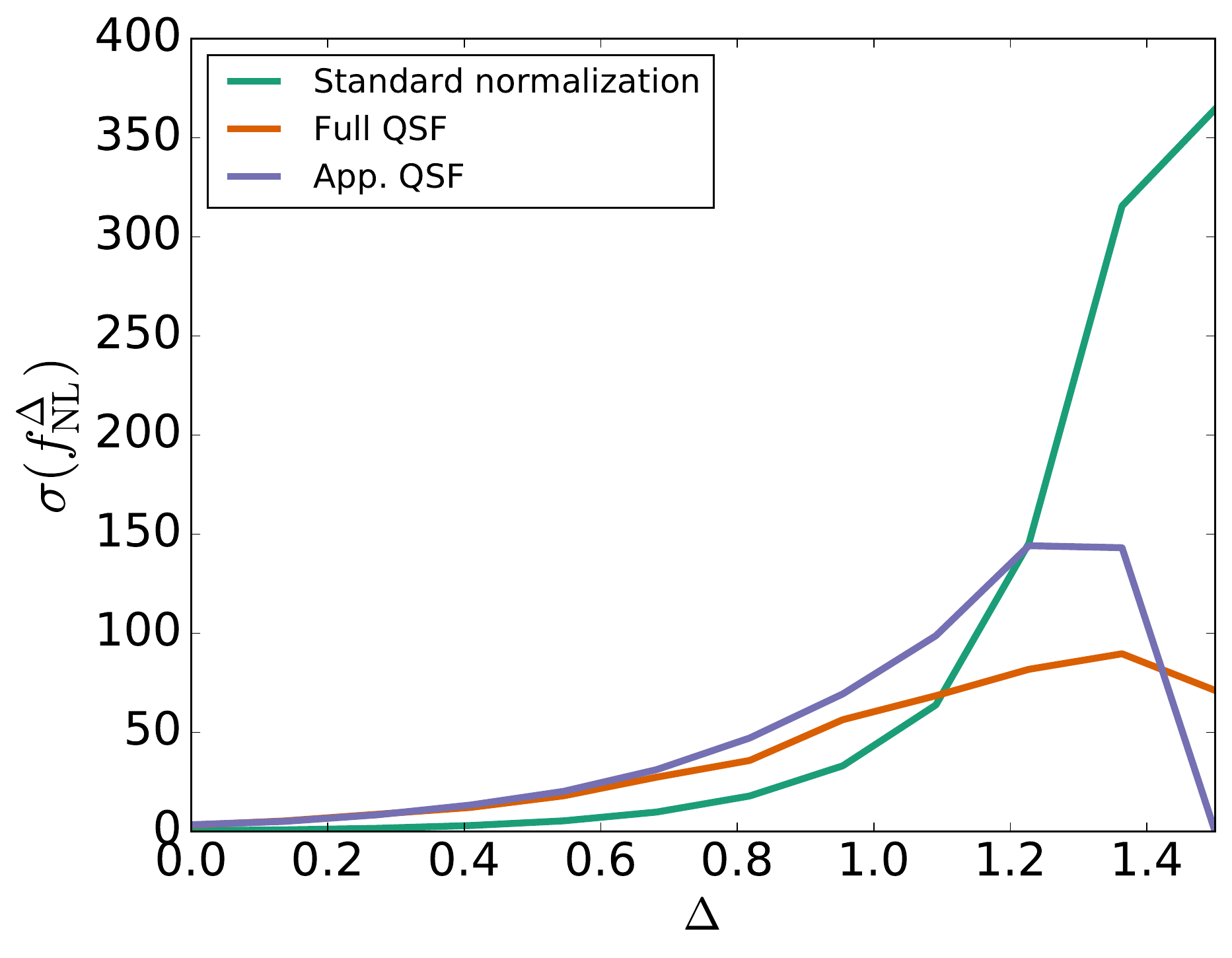}
\caption{Evolution of the constraint on $\fnl$ as a function of $\Delta$, for different normalizations of the scale-dependent bias $\bfnl$. The green curve corresponds to a scale-dependent bias given by eq.~\eqref{normEquiDelta}. The purple one is given by the local limit of the scale-dependent bias, Eq.~\eqref{QSFApp}, and is just a $\Delta$ dependent rescaling eq.~\eqref{normEquiDelta} (which should not be trusted for $\Delta\to 3/2$). Finally, the orange curve shows the result with the full QSF normalization of eq.~\eqref{eq:dbsd}. At low $\Delta$, where the PNG is dominated by the squeezed limit, the QSF normalization is well approximated by eq.~\eqref{QSFApp} and the constraint with standard normalization is about 6 times better. As we go to higher value of $\Delta$, the approximation eq.~\eqref{QSFApp} starts to fail: $\sigma(\fnl)$ goes to zero because the normalization diverges at $\Delta=3/2$, and should not be trusted. Moreover, the QSF normalization gets much better constraints than with the normalization eq.~\eqref{normEquiDelta}. This is because when using eq.~\eqref{eq:dbsd}, the signal is much less degenerate at high $k$: one can check that for $\Delta=3/2$ the signal is of the same order with eq.~\eqref{normEquiDelta} and eq.~\eqref{eq:dbsd}. This difference in the constraints explains why we cannot interpret Fig.~\ref{fig:QSFPlanck} with insights from Fig.~\ref{fig:DeltaVaryings}.}  
\label{fig:diffNorm}
\end{figure}

To be concrete, the transition in Fig.~\ref{fig:DeltaVaryings} occurs at $\Delta \sim 1$ and $\sigma(\fnl^{(\Delta)}) < 10$ for $\Delta < 0.5$.  A significant difference between these figures is the normalization of the squeezed limit relative to the equilateral limit.  For the purpose of a CMB experiment, all of the constraining power is coming from equilateral configurations.  For $\Delta \sim 0.5$ we find that the amplitude of the QSFI normalization is suppressed by a factor of 4-5 compared to the expectation from the amplitude in the equilateral configurations.  We compared in Fig.~\ref{fig:diffNorm} the effect of the different normalizations on $\sigma(\fnl^{\rm (\Delta)})$ for fixed values of $\Delta$.
One can see in Fig.~\ref{fig:diffNorm} that at low $\Delta$, where the signal gets a significant contribution from the local configuration, the approximation in eq.~\eqref{QSFApp} performs well. However, as the $\Delta$ increases and the signal shifts toward higher $k$, the approximation breaks down. Moreover, as $\Delta$ increase, even if one can check that the signal increase faster with the normalization of eq.~\eqref{normEquiDelta} than with the full QSF, the degeneracies are stronger with the former, resulting in a weakening of the constraint on $\fnl^{(\Delta)}$. 

The broad lesson here is that this relative normalization is important when comparing limits from scale-dependent bias with probes that are not limited to squeezed configurations.  This can be seen more easily if we imagine a model where the bispectrum is a sum of an equilateral and quasi-single field piece, $B = \fnl^{\rm comb, \Delta} ( B_{\rm eq}+ \alpha B_{\Delta})$.  Since scale-dependent bias cannot measure $\fnleq$ better than Planck, as $\alpha \to 0$, we will find that Planck places stronger constraints on $ \fnl^{\rm comb, \Delta}$ for all $\Delta$.  For QSFI, this suppression is significant and limits the improvement over Planck to $\Delta < 0.3$.  Given that $\fnl^{(\Delta)} < 20$ for $\Delta < 1$ (and that $\fnl^{(\Delta = 2)} = \fnleq$) it is not unreasonable to expect larger improvements could be available in other models with non-trivial scaling in the squeezed limit.  

\section{Multi-tracer technique}

\label{multi-tracer}

\subsection{The set-up}

Since our constraints depend on measuring the power spectrum, {\it a priori} they suffer from a fundamental limit, the so-called cosmic variance limit. This is just to say that even if we were perfectly sampling the density field, because we only have access to a finite number of modes (especially for long modes), there is an error that cannot be beat. While this is certainly true when using a single tracer, it was proposed in \cite{Seljak:2008xr} that by using more than one tracer, one can get a measurement of the bias (and therefore $\fnl$) that is not cosmic-variance limited. This has recently been applied to constraints on PNG from the galaxy bispectrum \cite{Yamauchi:2016wuc}.

Let's take the example of  two tracers, related to the density field $\delta$ through $\delta_1=b^{(1)}\delta$ and $\delta_2=b^{(2)}\delta$. While each separate measure of $\delta_i$ is limited by cosmic variance since one needs to know the stochastic variable $\delta$, taking the ratio is independent of the realization of the underlying field $\delta$. It depends only on the ratio of the biases, which we can thus measure without the cosmic-variance limit.

In order to see what type of improvement might be gained with this method, we also ran Monte Carlo Markov chains with two tracers keeping the same configuration (volume, redshift, etc) as before for the survey. Each one of them has its own independent set of bias parameters as in eq.~\eqref{PhhRen}, meaning that we have twice as many bias parameters as in the single-tracer case (but still a single $\fnl$ parameter). In this case, the quantity of interest is the matrix (the extension of the power spectrum in the single tracer case)
\be
\label{eq:matC}
 C=\left(
\begin{array}{cc}
P_{11} & P_{12}
\\
P_{12} & P_{22}
\end{array}
\right)\,,
\ee
where the $P_{ij}$ are the (cross) power spectra between the tracers $i$ and $j$. Then, by analogy with the single tracer case, the likelihood is
\be
\begin{split}
\log{\mathcal{L_{\rm Double}}(\hat{C}(k)|\btheta)}&=-\frac{N_k}2\left\{\Tr\left[\hat{C}(k)C_{\rm th}^{-1}(k,\btheta)\right]-\Tr\left[\log\left(\hat{C}(k)C_{\rm th}^{-1}(k,\btheta)\right)\right]\right\}\,,\\
&=-\frac{N_k}2\left\{\Tr\left[\hat{C}(k)C_{\rm th}^{-1}(k,\btheta)\right]-\log\left[\det \left(\hat{C}(k)C_{\rm th}^{-1}(k,\btheta)\right)\right]\right\}\,.
\end{split}
\ee
Here, $\hat{C}$ is the matrix \eqref{eq:matC} using the fiducial (cross) power spectra $P_{ij}\rightarrow \hat{P}_{ij}$ and $C_{\rm th}$ using the theory ones  $P_{ij}\rightarrow P_{{\rm th}, ij}$, given by the straightforward extension of eq.~\eqref{PhhRen}.

The decisive parameter here is the number density and the associated shot noise. When the density $\bar n$ is low, the shot noise $1/\bar n$ is high. This means that the total noise, given by $C_{\rm th}$, is dominated by the shot noise contribution -- remember that $P_{\rm th}$ is given by eq.~\eqref{PhhRen} -- and not by the part proportional to $P_{\rm G}$. In this case, the limiting factor is thus not cosmic variance, and using the multi-tracer technique cannot improve the constraints. However, as soon as the shot noise is low enough for $P_{\rm G}$ to dominate the noise, using the multi-tracer technique can dramatically improve the constraints on $\fnl$.
To illustrate this, we will take a toy model with two species, each one with a different value for its fiducial bias, which is a requirement for the multi-tracer approach to work (the improvement increases the more different the fiducial biases are \cite{Seljak:2008xr}). In this case the only thing varying will be the (fiducial) number density, which means that we will go from shot-noise dominated to cosmic-variance limited.

We will also study a case closer to actual surveys, where the first tracer comes from a galaxy population
with properties determined by a minimum halo mass, $M_{h,{\rm min}}$, analogously to the discussion around eq.~\eqref{biasmhmin}. We will not be free to change the number density, galaxy bias and Lagrangian radius independently, but instead they are jointly determined by $M_{h,{\rm min}}$. For the second tracer, we will directly take the matter density field $\delta$, that could in principle be obtained through weak lensing.

There is one small subtlety with $k_{\rm max}$. The Lagrangian radius $R_*$ depends on $M_{h,{\rm min}}$ and can get small for low minimum halo mass. To ensure that we stay in linear theory for our forecast, we impose that the cut-off scale $k_{\rm max}$ be the minimum between $1/(2 R_*)$ and the non linear scale, $\pi/(2R_{\rm NL})$, where $R_{\rm NL}$ is such that the r.m.s. of linear fluctuations in spheres of radius $R_{\rm NL}$ is $1/2$.

\subsection{The results}

\begin{figure}[h!]
\centering
\includegraphics[width=0.49\textwidth]{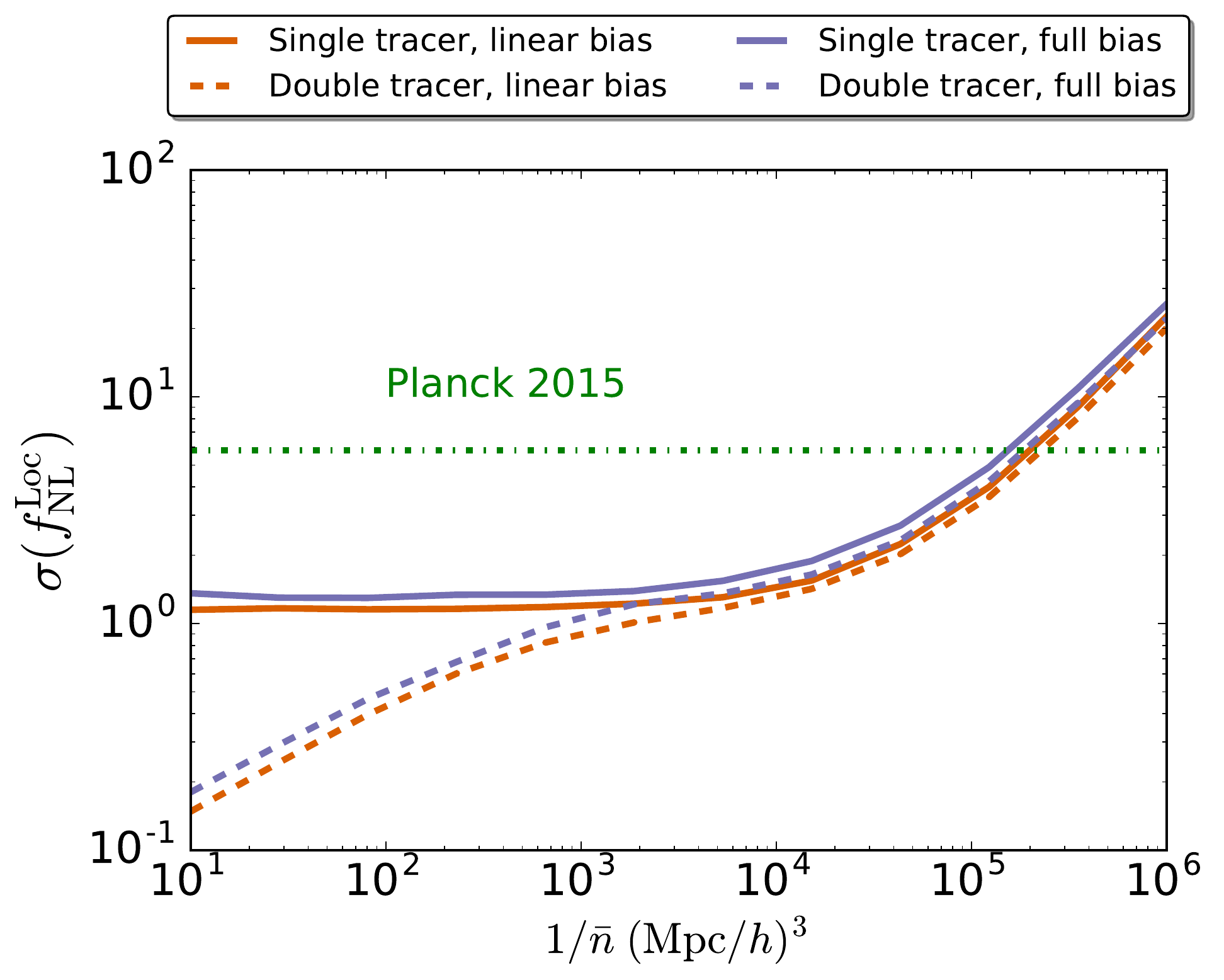}\;\includegraphics[width=0.49\textwidth]{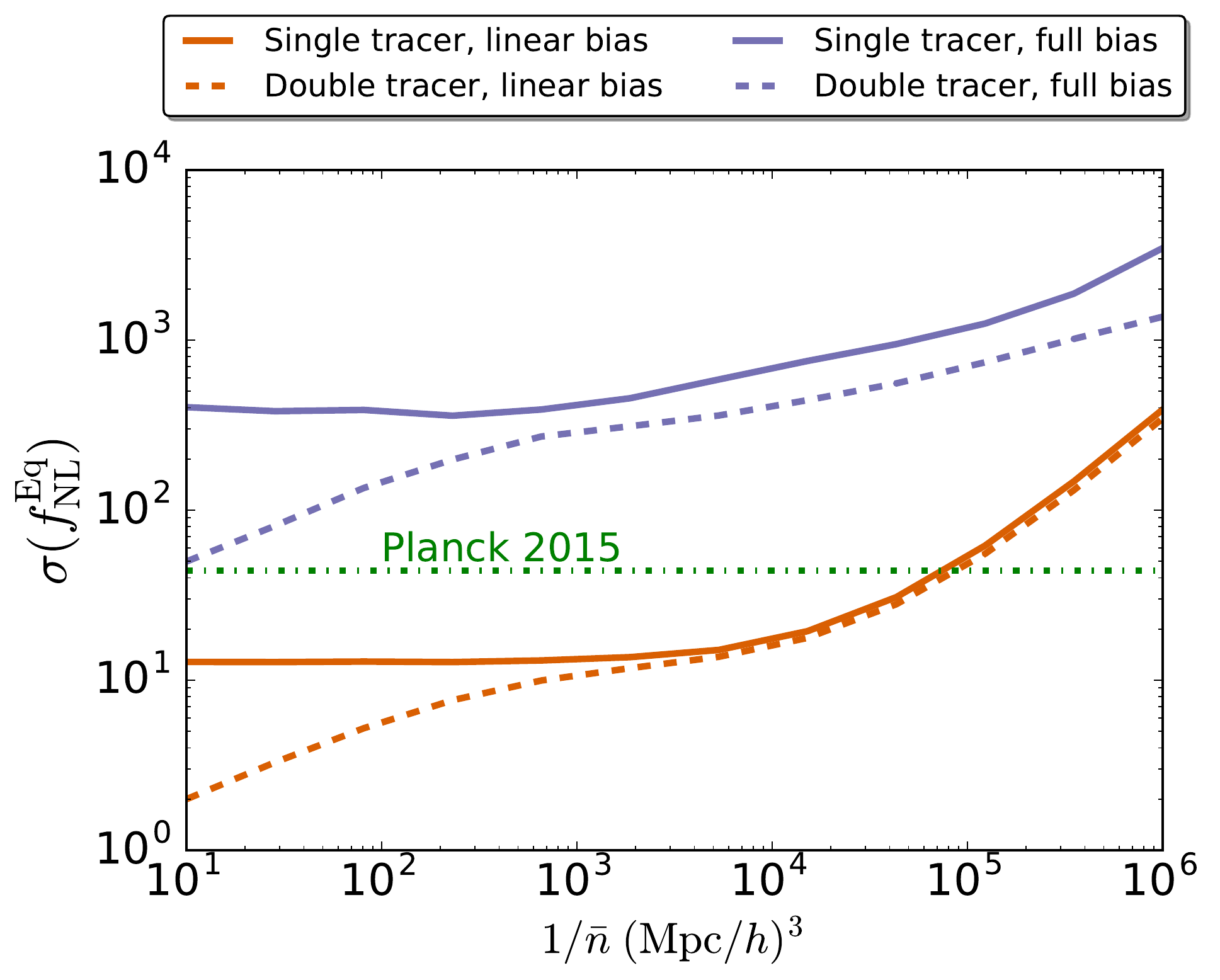}\\
\includegraphics[width=0.49\textwidth]{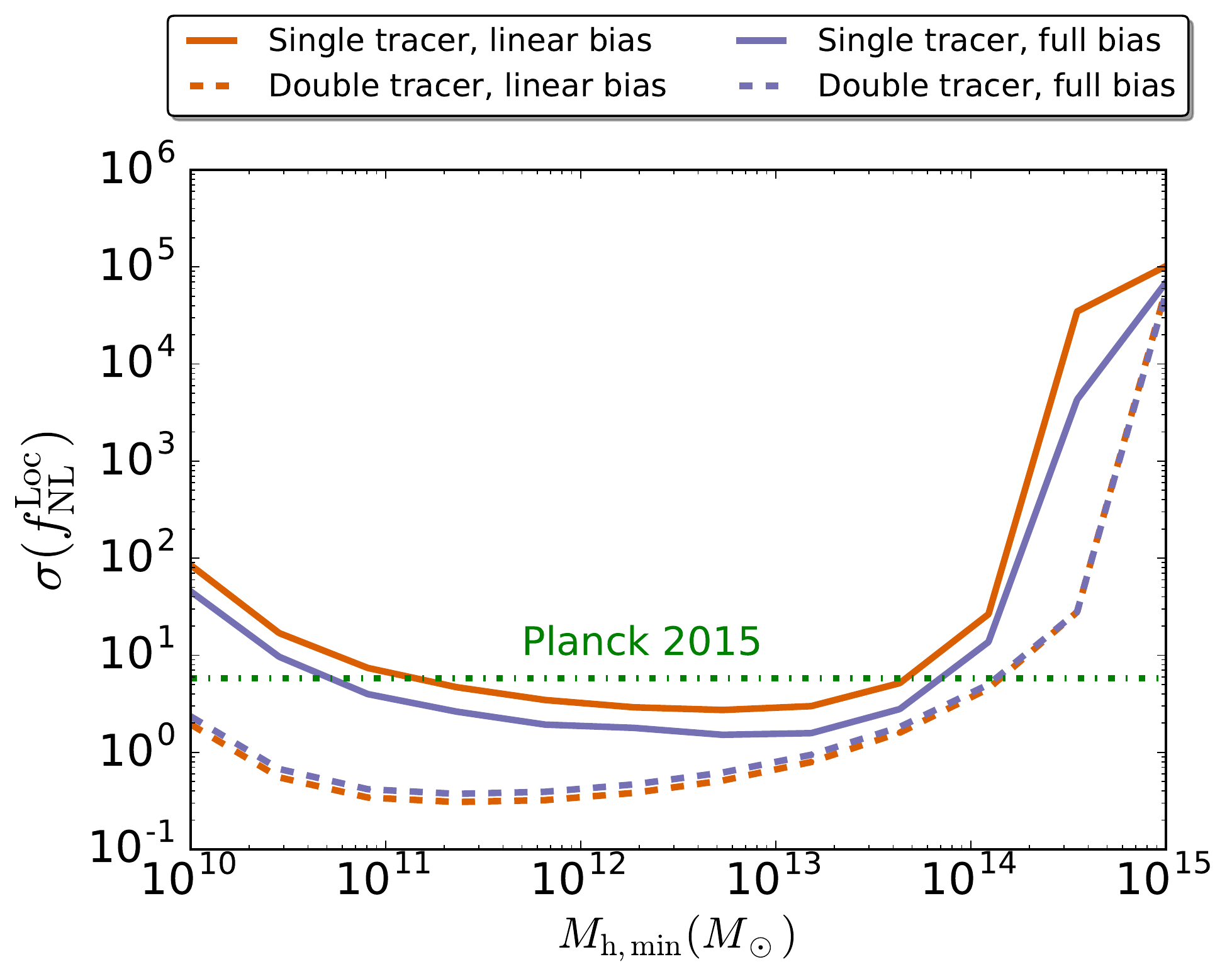}\;\includegraphics[width=0.49\textwidth]{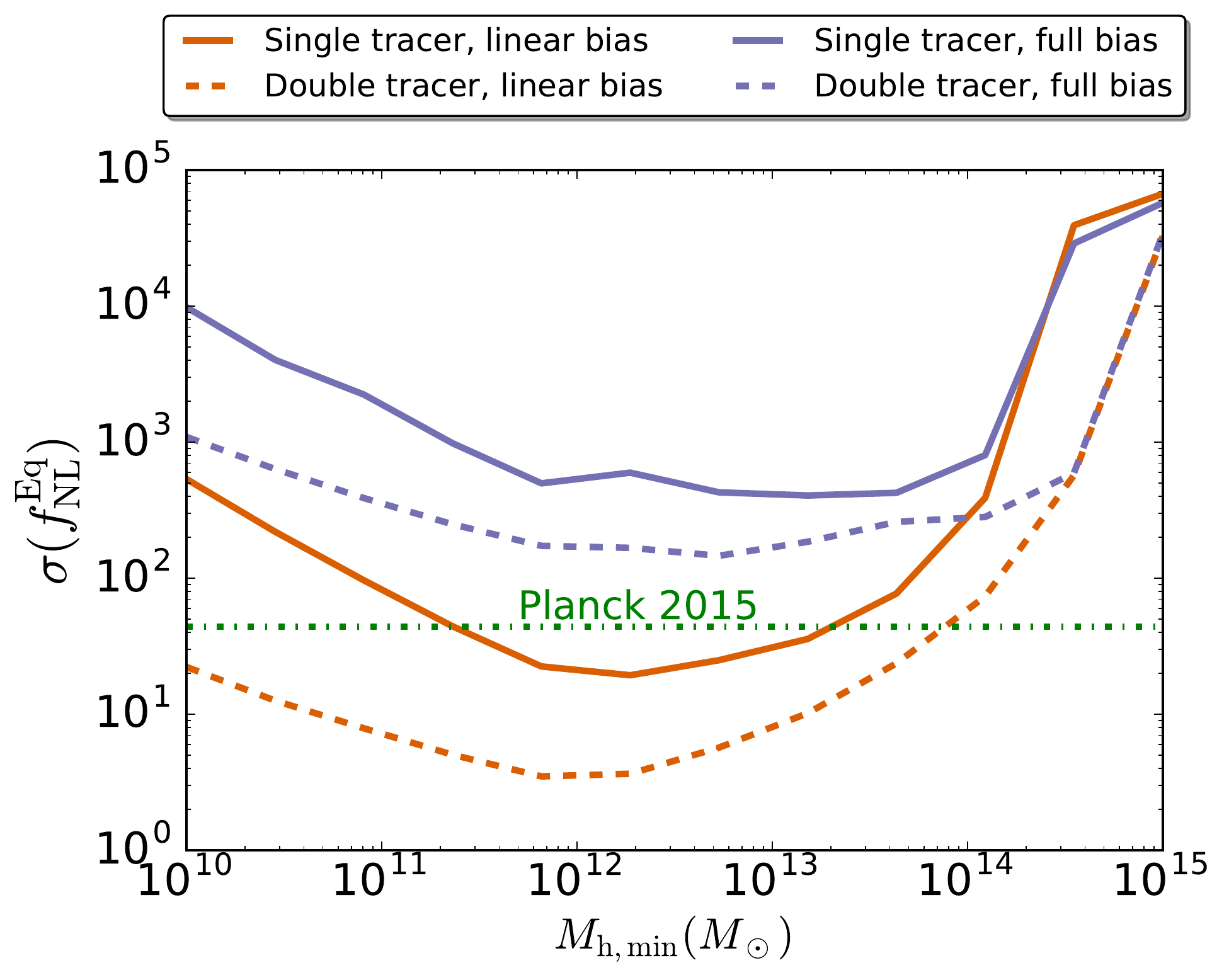}
\caption{\textit{Top}: Evolution of $\sigma(\fnlloc)$ (\textit{left}) and  $\sigma(\fnleq)$ (\textit{right}) as a function of the number density  for a single tracer (full lines) and two tracers (dashed lines). For the double-tracer case, we have two galaxy populations with different biases ($b_{\delta,1}=3.6$ and $b_{\delta,2}=2$) and same number density. The single tracer is the sum of those two populations, with bias $(b_{\delta,1}+b_{\delta,2})/2$ and twice the number density of the double-tracer case.  As the shot noise becomes negligible, the single-tracer case is limited by cosmic variance, while the double-tracer keeps on improving.  \textit{Bottom}: Evolution of  $\sigma(\fnlloc)$ (\textit{left}) and  $\sigma(\fnleq)$ (\textit{right}) as a function of the minimum halo mass  for a single tracer (solid lines) and two tracers (dashed lines). For the single tracer, the bias, Lagrangian radii and densities of galaxies $\bar{n}$ are computed from the halo mass function. For the double-tracer, the additional tracer is taken to be the dark matter field (possibly obtained by weak lensing). Using two tracers always constitutes an improvement by roughly an order of magnitude over a single one (see main text for details).
\newline In purple, the results when marginalized over all bias parameters, in orange when the marginalization is only over the linear bias(es). The green lines are the current limit from Planck \cite{Ade:2015ava}.}
\label{fig:multi-tracer}
\end{figure}

To illustrate how the multi-tracer approach allows to beat cosmic variance, we plot on the top panels of Fig.~\ref{fig:multi-tracer} the expected precision on $\fnlloc$ (\textit{left})  and $\fnleq$ (\textit{right}) as a function of the shot noise $1/\bar{n}$ (\textit{top}), where $\bar{n}$ is the number density of the population we consider. The two tracers are two different (idealized) population of galaxies, with different biases ($b_{\delta,1}=3.6$ and $b_{\delta,2}=2$) but same Lagrangian radius $R_*=3.7\, {\rm Mpc}/h$ and same number density. The single tracer is a combination of the two populations, with bias $(b_{\delta,1}+b_{\delta,2})/2$ and twice as many galaxies. While at low density (high noise) the behaviors are similar, when getting to larger densities, the single tracer sensitivity starts to plateau, limited by cosmic variance, while the double-tracer case keeps on improving.

To put this in a more realistic setting, we show in the bottom panels of Fig.~\ref{fig:multi-tracer} the expected precision on $\fnlloc$ (\textit{left}) and $\fnleq$ (\textit{right})  as a function of the minimum halo mass defining one of the tracers, where the other tracer is taken to be a direct measurement of the dark matter density.  The minimum halo mass determines the number density and mean bias of the sample in accordance with N-body simulations.

At low minimum halo masses, the density is higher, implying lower shot noise, and the multi-tracer approach is expected to improve the constraints. However, the bias gets closer to one, which both decreases the signal and the effect of the multi-tracer approach (which is more efficient the higher the difference in biases is \cite{Seljak:2008xr}). On the other hand, at high minimum halo mass, the bias is very different from one, but the density is much lower, which means the signal is dominated by shot noise and therefore the multi-tracer approach is not effective. 

Looking at the curves in the single tracer case (in purple), one can also see why we chose the default value of $M_{\rm h,min}=10^{13}M_\odot$, as it is the one the approximately minimizes $\sigma(\fnleq)$. 

On each panel of Fig.~\ref{fig:multi-tracer}, we have made a comparison between our MCMC forecasts (with a fiducial volume $V=100 \,(\text{Gpc}/h)^3$) and the current constraints from Planck \cite{Ade:2015ava}. While for local PNG, we can (optimistically) improve by an order of magnitude on the Planck value, it is harder to imagine reaching those level for the equilateral PNG. Even at the minimum, our constraints are still a factor $\sim3$ higher than those from Planck.

\begin{figure}[h!]
\centering
\includegraphics[width=0.65\textwidth]{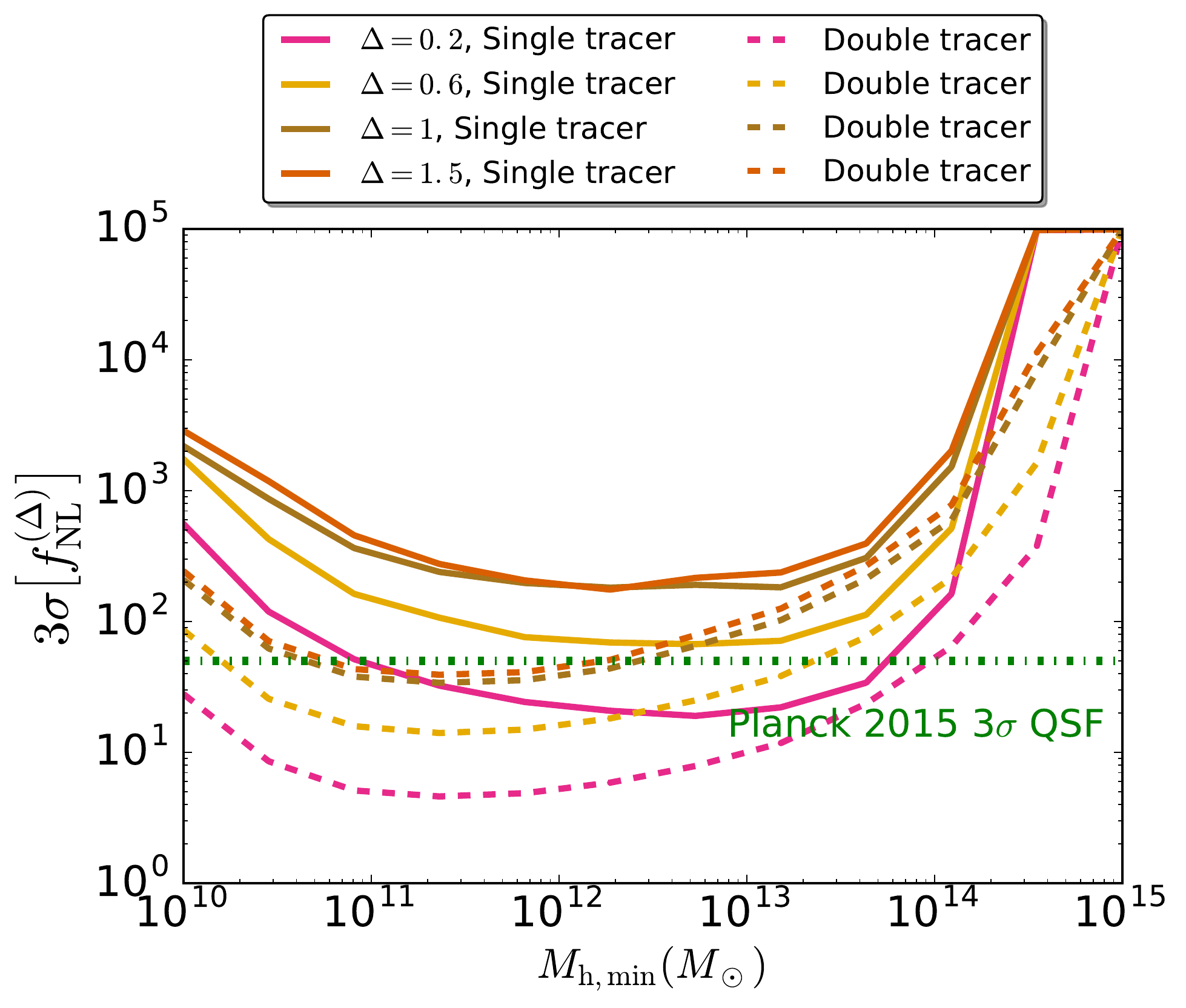}
\caption{The evolution of $99\%$ bound on $\fnl^{(\Delta)}$ from quasi-single field inflation (see Section \ref{sec:beyond}), as a function of the minimum halo mass of the galaxy sample. The solid lines are for an analysis with this sample only, while the dashed one combine it with a second tracer, dark matter. For all values of $\Delta$, there is an improvement over Planck when using two tracers and sufficiently small minimum halo masses (which means larger number density and smaller shot noise).}
\label{fig:multi-tracerQSFfix}
\end{figure}

\begin{figure}[h!]
\centering
\includegraphics[width=0.65\textwidth]{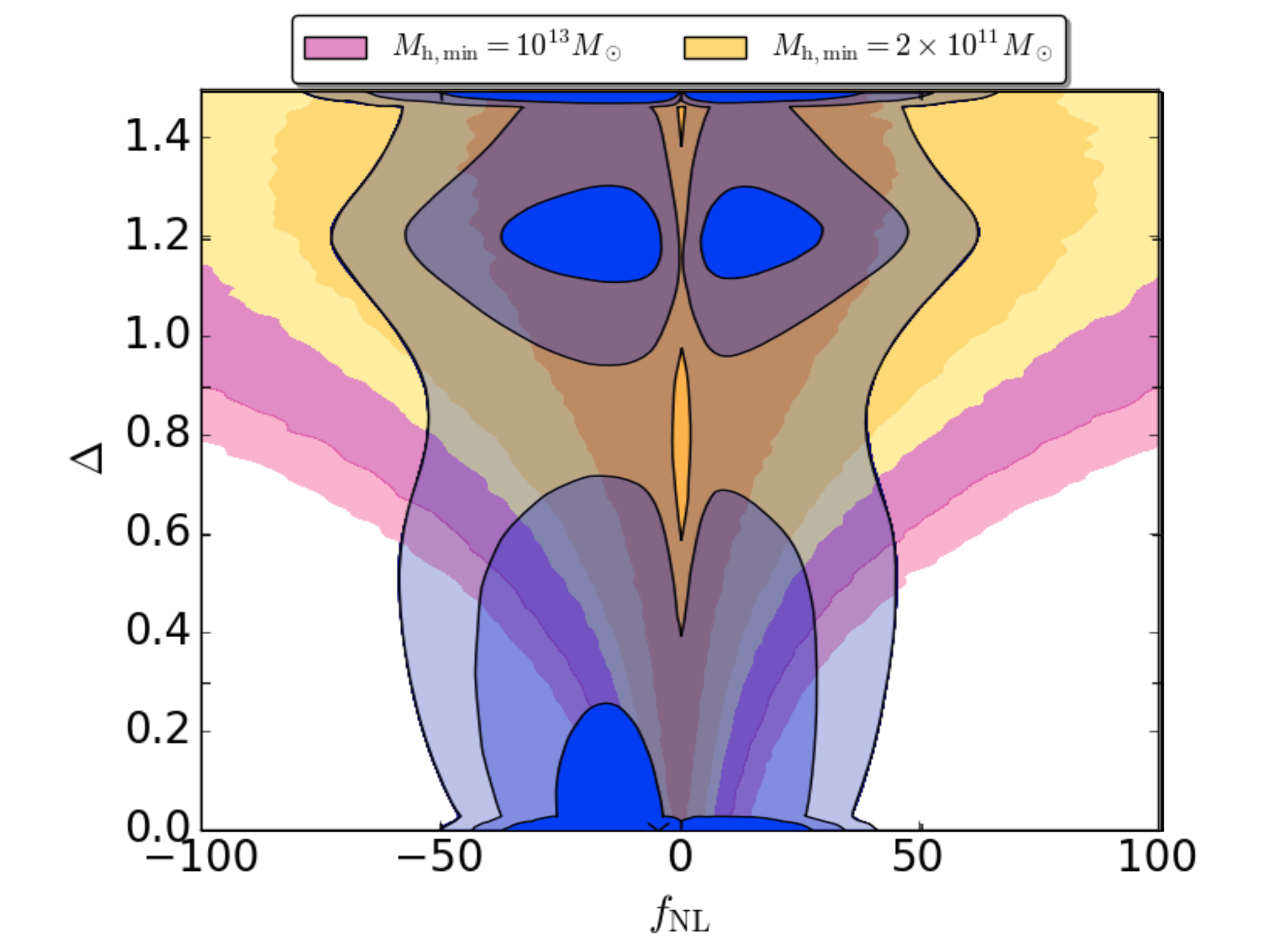}
\caption{The green regions are the $68\%$, $95\%$ and $99\%$ contours from our MCMC with two tracers, dark matter and a galaxy sample with $M_{\rm h,min}=10^{13} M_\odot$, similar to our analysis in the single tracer case. The red regions are the $68\%$, $95\%$ and $99\%$ contours when considering dark matter and a galaxy sample with $M_{\rm h,min}=2\times10^{11} M_\odot$. The blue are the $68\%$, $95\%$ and $99\%$ contours from Planck \cite{Ade:2015ava}. While for the green contours, we recover only a minor advantage over the single tracer case of Fig.~\ref{fig:QSFPlanck}, for lower minimum halo mass (and therefore lower shot noise), the multitracer approach provides an improvement over Planck for almost all values of $\Delta$. }
\label{fig:multi-tracerQSF}
\end{figure}

However, these plots show that there is no ``error floor" due to the degeneracies with other bias parameters that would prevent measuring $\fnleq$ using scale-dependent bias. Degeneracies are still an important factor in the poor performance, but this adds to the fact that the signal is weak compared to local PNG, or in other words that its signal peaks for bispectrum configurations away from the squeezed limit domain. We can conclude that equilateral non-Gaussianity is not as ideal for scale-dependent bias as the local type for both reasons above. Nevertheless, for the case of quasi-single field inflation described in Section \ref{sec:beyond}, the results are more promising, with potential improvement over Planck for most of the parameter space (see Figs.~\ref{fig:multi-tracerQSFfix} and \ref{fig:multi-tracerQSF}).

To gauge where one can expect an improvement in a setup similar to the bottom panels of Fig.~\ref{fig:multi-tracer}, we plot in Fig.~\ref{fig:multi-tracerQSFfix} the expected value of the $99\%$ bound on $\fnl^{(\Delta)}$, for fixed values of $\Delta$ (using the normalization of eq.~\eqref{eq:dbsd}. We compared it to an estimate of the bound from Planck, approximating the $3\sigma$ contours of Fig.~\ref{fig:QSFPlanck} by a constant value of $50$. $M_{\rm h,min}$ is the minimum halo mass of the galaxy sample that constitutes the first tracer, the second being dark matter.

At first sight, it seems that for any value of $\Delta$, one can find a range of minimum halo masses where double-tracer improves on Planck. However, the analysis in Fig.~\ref{fig:multi-tracerQSFfix} assumes a fix value of $\Delta$, while in Fig.~\ref{fig:QSFPlanck} it was allowed to vary, bringing potential new degeneracies. This is why, in order to compare with Fig.~\ref{fig:QSFPlanck}, we look at the constraints in the plane $(\fnl,\Delta)$ using two tracers, dark matter and galaxies. To see how the improvement changes with minimum halo mass as in Fig.~\ref{fig:multi-tracer}, we plot two set of contours for two different minimum halo masses. The results are shown in Fig.~\ref{fig:multi-tracerQSF}

When looking at doing the same analysis as before for galaxies (i.e. setting $M_{\rm h,min}=10^{13} M_\odot$) but using the information in the dark matter field as well, the region where scale dependent bias improves on Planck goes from $\Delta\lesssim 0.2$ to $\Delta \lesssim 0.6$. Moreover, as soon as we use more galaxies by setting $M_{\rm h,min}=2\times10^{11} M_\odot$, the shot noise reduces and the bounds are better than Planck's for more than half of the range in $\Delta$. In order to see that, one has to use the full scale-dependent bias eq.~\eqref{eq:dbsd} and not the approximate eq.~\eqref{QSFApp}. In particular, it is not necessarily easy to compare those results to the equilateral case, we showed in Fig.~\ref{fig:diffNorm} that the two normalization \eqref{normEquiDelta} and \eqref{eq:dbsd} give very different results for $\Delta \rightarrow 1.5$.

\section{Summary and Conclusions}\label{sec:conc}

The ongoing search for primordial non-Gaussainty will continue to sharpen our understanding of the very early history of the universe and the origin of structure.  Large-scale structure offers great potential for increasing our sensitivity over current limits from the CMB.  However, large-scale structure is non-Gaussian in its own right and separating the two effects will be challenging.

In the case of local non-Gaussianity, scale-dependent bias is a particularly compelling target as it is distinguishable  from effects of late time evolution and can be measured directly from the power spectra.  The signal arises from squeezed configurations of the primordial correlation functions (especially the three-point function), which are strongly constrained by symmetries \cite{Creminelli:2004yq}.  In the context of inflation, the squeezed limit is sensitive to the spectrum of particles and their interactions around the time of horizon crossing.
 
In this paper, we explored the use of scale-dependent bias for constraining non-Gaussianity beyond the local shape.
In general, the scale-dependent bias signal can be expected to be partially degenerate with contributions to the halo power spectrum related to non-linear evolution, in particular the non-linear evolution of matter fluctuations, non-linear biasing between halos and matter, and non-local halo biasing.
A major focus of this work was to include these contributions in a comprehensive manner and to study
primordial non-Gaussianity constraints after marginalization over this non-primordial physics.

We were particularly interested in the scale-dependence that arises from the transfer function between the {\it primordial} Newtonian potential (for which the mode-coupling due to primordial non-Gaussianity is specified) and the potential during matter domination.  In principle, this makes any primordial mode-coupling distinguishable from mode-coupling due to non-linear structure formation.  Here we quantified the degree to which equilateral and quasi-single field shapes can be measured in this way and studied the degeneracy with the parameters of a general biasing model.

For local and some quasi-single field shapes, significant improvements over Planck are achievable for a large volume large-scale structure survey: for example, $\sigma(\fnlloc)\simeq 1$ could reached (similar to the forecasts of future surveys in \cite{Carbone:2008iz,dePutter:2014lna,Dore:2014cca}) with a single tracer of the matter density (e.g.~a single galaxy sample), which could be improved by a factor of 5 with a multi-tracer approach. This is a level that is interesting from a theoretical point of view \cite{dePutter:2016trg}. For equilateral non-Gaussianity, after marginalizing over a
comprehensive halo biasing model, the sensitivity falls below the current Planck limits, reaching $\sigma(\fnleq)\simeq 500$ with a single tracer, and about $145$ with two tracers. The latter is about $3$ times worse than current Planck sensitivity.  Nevertheless, there is information encoded at large distances, where the transfer function is non-trivial, that can in principle be measured without cosmic variance.  For this reason, forecasting using the multi-tracer technique shows significant improvements for equilateral non-Gaussianity. Moreover, the case of quasi-single field inflation is even more promising, with significant improvement possible over a good portion of the parameter space.

Ultimately, a large-scale structure survey should also include the information in the galaxy bispectrum as part of its analysis of primordial non-Gaussianity.  For non-local shapes (with large signal in non-squeezed configurations), the galaxy bispectrum contains much more information than scale-dependent bias and could lead to significant improvements over Planck for all shapes because large-scale structure surveys have access to a much higher number of clustering modes than the CMB.  However, a galaxy bispectrum analysis presents a number of challenges that are not present for the power spectrum.  We have shown in this work that, optimistically, there is an interesting range of parameter space (beyond merely local-type non-Gaussianity) where improvements over Planck are achievable from the power spectrum alone.

\vspace{5mm}
\noindent
{\bf Acknowledgements}.
We would like to thank Fabian Schmidt for helpful discussions. Part of the research described in this paper was carried out at the Jet Propulsion Laboratory, California Institute of Technology, under a contract with the National Aeronautics and Space Administration. This research is partially supported by NASA ROSES ATP 14-ATP14- 0093 grant. RdP and O.D. acknowledge support by the Heising-Simons foundation.

\appendix
\section{Calculating the halo power spectrum}
\label{detailsPS}
By summing the contribution from all the diagrams described in Section \ref{subsec:nonlinnonloc}, one gets a halo-halo power spectrum of the following form
\beq
\label{Phhsimple}
\begin{split}
P_{hh}(k)=&b_{\rm full}^2\,\left[P_{\rm G}(k)+P_{22}(k)+P_{13}(k)\right]+P_{\epsilon_0}+2P_{\epsilon_{\delta}}P_{\rm G}(k)\\
+& 4 b_{\rm full}\left\{\int_\p\left[ b_{\delta^2}+ b_{s^2}\left(\mu_{-}^2-\frac13\right)\right]\left[P_{\rm G}(|\q-\p|)-P_{\rm G}(p)\right]P_{\rm G}(p)\right\}\\
+&2 \int_\p\left[b_{\delta^2}+ b_{s^2}\left(\mu_{-}^2-\frac13\right)\right]^2P_{\rm G}(|\q-\p|)P_{\rm G}(p)\\
+&b_{\rm full}\, P_{\rm G}(k)\bigg\{6 b_{\Pi \Pi^{(2)}}\,\int_\p\frac27\left[\left(\frac{\k\cdot\p}{k\,p}\right)^2-1\right]\frac{2\mu^2}3P_{\rm G}(p)\\
&\quad+8 \, b_{s^2}\int_\p\left(\mu_{-}^2-\frac13\right)F_2(-\p,\k)P_{\rm G}(p)+8\, b_{\delta^2}\int_\p F_2(-\p,\k)P_{\rm G}(p)\bigg\}\,.
\end{split}
\eeq
We have used
\beq
\mu_{-}\equiv \frac{(\k-\p)\cdot\p}{|\k-\p|\,p}\,,
\eeq
we recall that (see standard cosmological perturbation theory \cite{Bernardeau:2001qr})
\beq
F_2(\k_1,\k_2)= \frac57+\frac27 \left(\frac{\k_1\cdot\k_2}{k_1\,k_2}\right)^2+\frac12\,\frac{\k_1\cdot\k_2}{\,k_1\,k_2}\left[\frac{k_1}{k_2}+\frac{k_2}{k_1}\right]\,,
\eeq
and $P_{\rm G}$ is the power spectrum of $\delta_{\rm G}$. Let us go through the various terms in this lengthy expression

\begin{itemize}
\item The first line is the extension of SPT result including $b_{\rm full}$ plus the contribution from stochastic fields. The last one, coming from $\epsilon_\delta \delta$, can be reabsorbed in the definition of the linear bias $b_\delta$.

\item The second line corresponds to Type 1) diagrams of Fig.~\ref{fig:Diagrams}, where one of the kernel is $F_2$. The factor 4 in front of the first curly parentheses comes for symmetry reasons: looking at the left of Fig.~\ref{fig:Diagrams}, there are two diagrams of the sort depending on how you connect the lines, and the extra factor of two comes for changing $G$ and $G'$. 

\item The third line corresponds to Type 1) diagrams of Fig.~\ref{fig:Diagrams}, where both kernels are different from $F_2$. The factor 2 comes again for symmetry reasons.

\item The third line is the only contributing diagram of Type 2) that is not SPT. The factor of 6 comes from the symmetries of the diagram, the factors $2/3$ and $2/7$ from the definition of the kernel.
\item  In the last line we find the diagrams of the third type. 
\end{itemize}

Once we have this expression in terms of ``bare" quantities, we can remove the parts that lead to divergences\footnote{Some of them only appear for $k\rightarrow 0$, which can be removed by replacing $P_{\rm G}(|\k-\p|)\rightarrow P_{\rm G}(|\k-\p|)-P_{\rm G}(p)$.}  as well as perform angular integration to simplify it. Morevoer, the last term of the first line can be reabsorbed in a redefinition of $b_\delta$, and will thus disappear from the expression.  We then obtain

\beq
\label{PhhRen2}
\begin{split}
P_{hh}(k)=&b_{\rm full}^2\,\left[P_{\rm G}(k)+P_{22}(k)+P_{13}(k)\right]+P_{\epsilon_0}\\
+& 4 b_{\rm full}\left\{\int_\p\left[ b_{\delta^2}+ b_{s^2}\left(\mu_{-}^2-\frac13\right)\right]\left[P_{\rm G}(|\q-\p|)-P_{\rm G}(p)\right]P_{\rm G}(p)\right\}\\
+&2 \int_\p\left[b_{\delta^2}+ b_{s^2}\left(\mu_{-}^2-\frac13\right)\right]^2\left[P_{\rm G}(|\k-\p|)-P_{\rm G}(p)\right]P_{\rm G}(p)\\
+&b_{\rm full}\,P_{\rm G}(k)\,\bigg\{6 b_{\Pi \Pi^{(2)}}\,\int_\p\left(\frac27\left[\left(\frac{\k\cdot\p}{k\,p}\right)^2-1\right]\frac{2\mu_{-}^2}3+\frac8{63}\right)P_{\rm G}(p)\\
&\qquad+8 \, b_{s^2}\int_\p\left[\left(\mu_{-}^2-\frac13\right)F_2(-\p,\k)-\frac{34}{63}\right]P_{\rm G}(p)\bigg\}\,.
\end{split}
\eeq
In the last two terms, the integration over the angle can be done analytically, and one finds that the contributions are proportional. Therefore, we can regroup them into a single one, keeping only the $b_{\Pi \Pi^{(2)}}$ term, leading to eq.~\eqref{PhhRen}.

\bibliographystyle{utphys}
\bibliography{refs}

\end{document}